\documentclass[amsmath,amssymb,aps,pre,twocolumn,superscriptaddress]{revtex4-2} 
\usepackage{amsmath,amsfonts}
\usepackage{soul}
\usepackage{dsfont}
\usepackage{graphicx}
\usepackage{bm}
\usepackage{hyperref}
\usepackage{url}
\usepackage[english]{babel}
\usepackage[utf8x]{inputenc}
\usepackage[T1]{fontenc}

\usepackage{color}
\definecolor{darkgreen}{rgb}{0,0.6,0}
\definecolor{darkblue}{rgb}{0,0,0.6}
\definecolor{darkred}{rgb}{0.6,0,0}
\definecolor{darkpurple}{rgb}{0.5,0,0.5}

\begin{document}

\title{Three-dimensional formulation of curved nematic shells}

\author{Mathieu Dedenon}
\email{Mathieu.Dedenon@unige.ch}
\affiliation{Department of Biochemistry, University of Geneva, 1211 Geneva, Switzerland}
\affiliation{Department of Theoretical Physics, University of Geneva, 1211 Geneva, Switzerland}


\date{\today}
\begin{abstract}
In soft matter, the phase of nematic liquid crystals can be made from anisotropic molecules in single component materials, or as a suspension of mesoscopic nematogens. The later offers more versatility in the experimental design of complex shapes, in particular thin curved shells, and is often found in biological systems at multiple scales from cells to tissues. Here, we investigate theoretically the transition from three-dimensional nematics to a two-dimensional description restricted to a tangent plane, using a mean-field approach. We identify a transition from first to second order isotropic-nematic transition in presence of weak tangential anchoring. Then, we clarify the conditions under which a two-dimensional description of thin nematic shells is relevant. Nonetheless, using the example of active nematic stress, we identify physical differences between two- and three-dimensional descriptions in curved geometry. Finally, we construct a thin film approximation of nematohydrodynamics for a nematic shell in contact with a curved substrate. All together, those results show that a tangential restriction of nematic orientation must be used with care in presence of curvature, activity, or weak anchoring boundary conditions. 
\end{abstract}

\maketitle

\section{Introduction}

Liquid crystals are fluids with orientational order emerging from anisotropic properties at the molecular level. The simplest example of such anisotropic fluids is an assembly of elongated molecules with head-tail symmetry, forming a nematic phase at sufficiently low temperature~\cite{gennes1993}, or high density for mesoscopic suspensions~\cite{Onsager1949}. Without externally-induced orientation, the principal direction of the nematic phase is chosen randomly, corresponding to spontaneous symmetry breaking. Nematic liquid crystals form an important class of materials with technological relevance, because their orientation is easily controlled by external fields or boundary conditions. Hence, the influence of anchoring boundary conditions has been well studied for three-dimensional samples of nematic liquid crystals~\cite{gennes1993}.

The study of thin nematic shells, where one sample dimension is much smaller than the two others, has been recently motivated by several experimental cases.
In the context of colloidal interactions, the rich phenomenology of closed nematic shells has been found to be sensitive to the layer thickness~\cite{FernandezNieves2007,Bates2010,LopezLeon2011}. In the context of active matter and biological physics~\cite{Marchetti2013,Ladoux2022b}, the local production of mechanical work around cytoskeletal filaments induces complex dynamics of the nematic texture when confined at a water-oil droplet interface~\cite{Sanchez2012,Keber2014,Giomi2017,Pearce2019}. At larger length scales, cell tissues cultured in vitro can form nematic layers with complex spatio-temporal patterns depending on mechanical activity~\cite{Gruler2000,Duclos2014,Sano2017,Duclos2018}. In the broader context of biology, some animals like \textit{Hydra vulgaris} are formed of thin nematic layers where topological defects have been linked to functional regions~\cite{MaroudasSacks2020,Ravichandran2025}.

Depending on sample preparation or boundary conditions, a nematic phase can exhibit regions with singularities in orientation called disclinations. In the limit of vanishing thickness for a thin nematic shell, corresponding to nematic surfaces, disclinations are reduced to point topological defects. Nematic surfaces exhibit specific properties like topological constraints on the defect-free configuration according to the Poincar\'e-Hopf theorem~\cite{Vitelli2006}, the coupling of topological defect charges with the Gaussian curvature~\cite{Vitelli2004}, or shape instabilities~\cite{Biscari2006,Frank2008}.

These examples motivate the theoretical study of thin nematic shells, or nematic surfaces in the limit of vanishing thickness, both for passive and active systems. The theoretical description of nematic shells is more complex than bulk nematics because of the coupling of nematic orientation to the geometry and topology of the shell. In general, the orientation of a nematic material is described at mean-field level by a second-rank tensor $\mathbf Q(\mathbf{r})$ field at position $\mathbf{r}$ in three-dimensional space. It captures the principal axis of orientation, called the director $\hat{\mathbf n}(\mathbf{r})$, and the dispersion of the nematogen orientational distribution through the scalar nematic order $S(\mathbf r)$ \cite{gennes1993}.
In the limit of infinitely thin nematic shells described by a surface $\mathcal S$, the orientations can be projected on the tangent planes of $\mathcal S$ and described with a two-dimensional tensor $\mathbf{Q}_{\mathcal S}$. In addition, when nematic order variations are negligible, the mean-field orientation is captured by a surface director field $\hat{\mathbf{n}}_{\mathcal S}$.
Some theoretical works used the director $\hat{\mathbf{n}}_{\mathcal S}$~\cite{Vitelli2004,Vitelli2006,Frank2008,Napoli2012a,Napoli2020,AlIzzi2023}, the two-dimensional tensor $\mathbf{Q}_{\mathcal S}$~\cite{Biscari2006,Virga2011,Virga2015,Pearce2019,Salbreux2022,Vafa2022,Mesarec2023,Wang2023}, or the three-dimensional tensor $\mathbf Q$ with a thin film approach~\cite{Napoli2012b,Napoli2013,Canevari2014,Golovaty2015,Golovaty2017,Novack2018,Voigt2018,Voigt2020} and direct simulations~\cite{Zhang2016,Giomi2023}.

The projection of orientations on a nematic surface $\mathcal S$ with the tensor $\mathbf{Q}_{\mathcal S}$ assumes perfect tangential anchoring of the nematogens~\cite{Biscari2006,Virga2011,Virga2015,Pearce2019,Salbreux2022,Vafa2022,Mesarec2023,Wang2023}.
In contrast, a soft tangential anchoring energy to the surface requires the three-dimensional description $\mathbf Q$, because of out-of-plane orientational fluctuations. Considering a soft tangential anchoring energy in the limit of high anchoring stiffness shows that a switch from $\mathbf Q$- to $\mathbf{Q}_{\mathcal S}$-descriptions is not trivial.
Importantly, the choice of $\mathbf{Q}_{\mathcal S}$ or $\mathbf Q$ formulations has physical consequences. It is well-known that the isotropic-nematic transition can be first order for bulk nematics~\cite{gennes1993}, allowing for phase coexistence, whereas it is of second order in two dimensions at mean-field level~\cite{Biscari2006,Virga2011,Voigt2018}.

We show in this article that thin nematic shells are fully described only with a three-dimensional formalism. We clarify the various approaches used in the literature, noting that a two-dimensional description can introduce physical artefacts in presence of nematic gradients. We show that the critical point from first-to-second order phase transition exists even for infinitely thin surfaces depending on the tangential anchoring strength.
Finally, we generalize the tensor theory of thin nematic shells from~\cite{Napoli2012b} by allowing for asymmetric tangential anchoring, which has direct applications in biological physics.

\begin{figure}
\centering
\includegraphics[width=1.\linewidth]{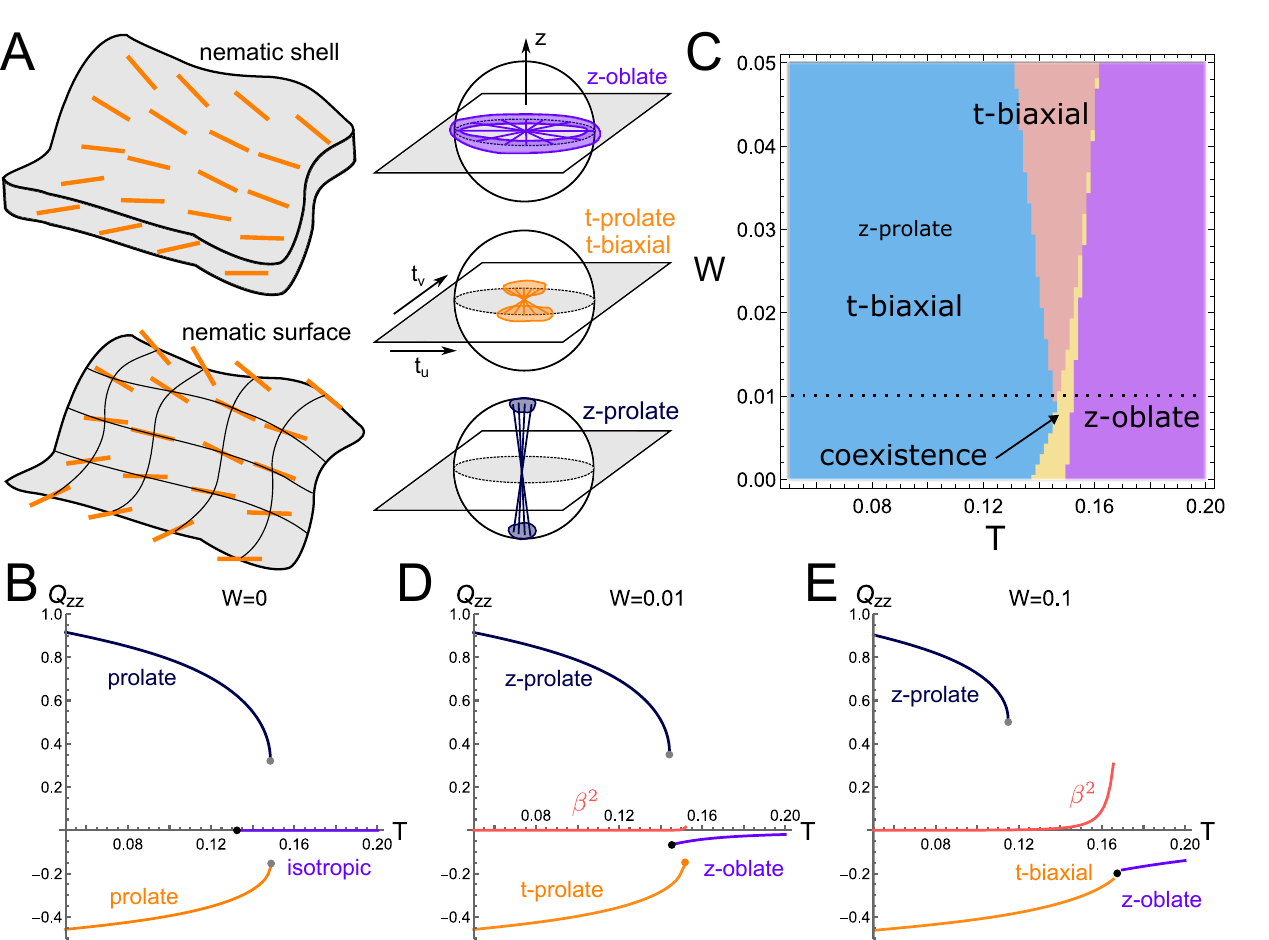}
\caption{Phase coexistence in the Maier-Saupe framework with soft tangential anchoring.
(\textbf{A}) Sketch of a thin nematic shell (top left) and a nematic surface (bottom left) in the limit of vanishing thickness. (Orange lines) nematic director. Distributions of $\hat{\mathbf u}$-orientations at a surface point: planar-isotropic called z-oblate (top right), planar-biaxial called t-biaxial (middle right), planar-uniaxial called t-prolate (middle right) and normal-uniaxial called z-prolate (bottom right).
(\textbf{B-E}) Equilibrium states of the Maier-Saupe potential with quadratic tangential anchoring.
Normal nematic tensor component $Q_{zz}$ as a function of $T$ for $W=0$ (B), $W=0.01$ (D), $W=0.1$ (E). (Red) biaxial coefficient $\beta^2$. (\textbf{C}) Phase diagram of the isotropic-nematic states in the parametric plane $(T,W)$. (Dashed line) case $W=0.01$.
}
\label{fig1}
\end{figure}

\section{Two- to three-dimensional tensorial description}
In this section, we present the different descriptions used for nematic orientational order at the mean-field level, in bulk and on a surface.
\subsection{Three-dimensional Q-tensor}
We start with the statistical definition of the three-dimensional order parameter $\mathbf Q$. For rod-like nematogens described by a single individual orientation vector $\hat{\mathbf u}$, the nematic tensor is a traceless symmetric tensor defined as
\begin{equation}\label{eq:q3d}
\mathbf Q=\frac{1}{2}(3\langle\hat{\mathbf u}\hat{\mathbf u}\rangle-\bm 1)
\end{equation}
where $\bm 1$ is the three-dimensional identity and average is made over a distribution of orientations $\rho(\hat{\mathbf u})$. An isotropic phase corresponds to $\mathbf Q=\bm 0$, such that $\langle u_i^2\rangle=1/3$ for any $i=[x,y,z]$ in Cartesian coordinates.

The most general configuration admits two orthogonal principal directions with $\mathbf Q=\frac{1}{2}s_1(3\hat{\mathbf n}_1\hat{\mathbf n}_1-\bm 1)+\frac{1}{2}s_2(3\hat{\mathbf n}_2\hat{\mathbf n}_2-\bm 1)$ and the nematic phase is called biaxial when the three eigenvalues of $\mathbf Q=\frac{1}{2}\mathrm{Diag}[2s_1-s_2,2s_2-s_1,-s_1-s_2]$ are distinct. It reflects the fact that the distribution of orientations is fully asymmetric, i.e. $\langle(\hat{\mathbf u}\cdot\hat{\mathbf n}_1)^2\rangle\neq\langle(\hat{\mathbf u}\cdot\hat{\mathbf n}_2)^2\rangle\neq\langle(\hat{\mathbf u}\cdot\hat{\mathbf n}_3)^2\rangle$ with $\hat{\mathbf n}_3=\hat{\mathbf n}_1\times\hat{\mathbf n}_2$.

The nematic order parameter $S$ is usually defined as the eigenvalue of $\mathbf Q$ with highest absolute value, and the associated eigenvector $\hat{\mathbf n}$ is called the director. From Eq.~\ref{eq:q3d}, one can show that $-1/2\leq S\leq 1$. When two directions have equivalent statistics, one talks about a uniaxial nematic phase and a diagonalization provides $\mathbf Q=\frac{S}{2}[3\hat{\mathbf n}\hat{\mathbf n}-\bm 1]$. The case $S>0$ called prolate corresponds to nematogens aligned along the axis of the director $\hat{\mathbf n}$. The case $S<0$ is called oblate because nematogens are essentially perpendicular to the director, see Fig.~\ref{fig1}A. Hence the director represents the direction of highest anisotropy. For instance, $S=-1/2$ corresponds to a planar-isotropic distribution with $\langle u_x^2\rangle=\langle u_y^2\rangle=1/2$ and $u_z=0$ for $\hat{\mathbf n}=\hat{\mathbf e}_z$, such that $\mathbf Q_{\rm pi}=1/4(\bm 1-\hat{\mathbf n}\hat{\mathbf n})-1/2\hat{\mathbf n}\hat{\mathbf n}$. The principal direction of orientation of the nematogens is given by the eigenvector associated to the highest positive eigenvalue of $\mathbf Q$.

\subsection{Two-dimensional Q-tensor}
Over a surface $\mathcal S$ with normal vector $\hat{\bm\nu}$, the restriction of orientation $\hat{\mathbf u}$ to the tangent plane is ${\mathbf u}_{\mathcal S}=\hat{\mathbf u}-u_{\nu}\hat{\bm\nu}=\bm 1_{\mathcal S}\cdot\hat{\mathbf u}$ with the surface projector $\bm 1_{\mathcal S}=\bm 1-\hat{\bm\nu}\hat{\bm\nu}$. Then, the statistical definition of a two-dimensional order parameter $\mathbf Q_{\mathcal S}$ is
\begin{align}\label{eq:q2d}
\mathbf Q_{\mathcal S}&=2\langle{\mathbf u}_{\mathcal S}{\mathbf u}_{\mathcal S}\rangle-\bm 1_{\mathcal S}
\end{align}
Only for purely tangential orientations, i.e. $u_{\nu}=0$, the tensor $\mathbf Q_{\mathcal S}$ is traceless and admits two eigenvalues with the same magnitude and opposite signs such that $\mathbf Q_{\mathcal S}=S_{2D}(2\hat{\mathbf n}_{\mathcal S}\hat{\mathbf n}_{\mathcal S}-\bm 1_{\mathcal S})$. In that case, the vector $\hat{\mathbf n}_{\mathcal S}$ is the surface director defined as the eigenvector with largest positive eigenvalue, and $S_{2D}=\hat{\mathbf n}_{\mathcal S}\cdot\mathbf Q_{\mathcal S}\cdot\hat{\mathbf n}_{\mathcal S}=\sqrt{\mathbf Q_{\mathcal S}:\mathbf Q_{\mathcal S}/2}$ is the surface nematic order satisfying $0<S_{2D}<1$. The planar-isotropic state with $\langle u_u^2\rangle=\langle u_v^2\rangle=1/2$ and $u_{\nu}=0$ gives $\mathbf Q_{\mathcal S}=0$, as expected.

\subsection{Tangential three-dimensional Q-tensor}
For thin nematic shells, the connection between the two descriptions is done by constructing level sets from a given surface $\mathcal S$ using normal coordinates along $\hat{\bm\nu}$, as shown in \cite{Napoli2012b}. For any surface $\mathcal S$, one can decompose the rod orientation as $\hat{\mathbf u}=\mathbf u_{\mathcal S}+u_{\nu}\hat{\bm\nu}$ and the identity tensor as $\bm 1=\bm 1_{\mathcal S}+\hat{\bm\nu}\hat{\bm\nu}$ within the orthonormal basis set $(\hat{\mathbf t}_u,\hat{\mathbf t}_v,\hat{\bm\nu})$.
Using Eq.~\ref{eq:q3d}, a tangential restriction of orientations $u_{\nu}=0$ gives
\begin{align}\label{eq-Qt}
\mathbf Q_{\rm tangent}
&=\frac{3}{4}\mathbf Q_{\mathcal S}+\mathbf Q_{\rm pi} \\ \nonumber
&=\frac{3}{2}\left[S_{2D}\hat{\mathbf n}_{\mathcal S}\hat{\mathbf n}_{\mathcal S}+\frac{1}{2}(1-S_{2D})\bm 1_{\mathcal S}-\frac{1}{3}\bm 1\right]
\end{align}
The first line of $\mathbf Q_{\rm tangent}$ is written as a sum of the surface tensor $(3/4)\mathbf Q_{\mathcal S}$ and the planar-isotropic tensor $\mathbf Q_{\rm pi}=\frac{1}{4}(\bm 1_{\mathcal S}-2\hat{\bm\nu}\hat{\bm\nu})$, such that $\mathbf Q_{\rm tangent}=\mathbf Q_{\rm pi}$ when $S_{2D}=0$. Using the director decomposition $\bm 1=\hat{\mathbf n}_{\mathcal S}\hat{\mathbf n}_{\mathcal S}+\hat{\mathbf n}_{\mathcal S,\perp}\hat{\mathbf n}_{\mathcal S,\perp}+\hat{\bm\nu}\hat{\bm\nu}$, one finds three distinct eigenvalues $S_n=(1+3S_{2D})/4$, $S_{n,\perp}=(1-3S_{2D})/4$ and $S_{\nu}=-1/2$. It shows that tangential anchoring corresponds to a biaxial state in the three-dimensional theory. This is easily understood from the restriction of fluctuations in the normal direction $\langle u_{\nu}^2\rangle=0$, Fig.\ref{fig1}A. A uniaxial state is recovered when $S_{2D}=0$ or $S_{2D}=1$.

The formulation of Eq.~\ref{eq-Qt} is equivalent to the approach of~\cite{Napoli2012b,Napoli2013} assuming perfect tangential anchoring.
However in Ref.~\cite{Fournier2005}, instead of using the statistical definition of nematic order, Eq.~\ref{eq:q3d}, the target state of their anchoring free energy is assumed to be uniaxial, hence it is equivalent to Eq.~\ref{eq-Qt} only for $S_{2D}=1$.

\section{First-to-second order Isotropic-to-nematic transition}
In this section, we first summarize the results of mean-field descriptions for the isotropic-nematic transition in bulk. We then add an anchoring energy to penalize out-of-plane orientations and study the parametric dependence on the nature of the transition.

\subsection{Landau-de Gennes free energy}
The nematic-isotropic transition in three dimensions is phenomenologically described using the Landau-de Gennes free energy density~\cite{gennes1993}
\begin{equation}\label{eq:fldg}
f_{LdG}(\mathbf Q)=\frac{a}{2}\mathrm{Tr}[\mathbf Q^2]-\frac{b}{3}\mathrm{Tr}[\mathbf Q^3]+\frac{1}{4}(\mathrm{Tr}[\mathbf Q^2])^2
\end{equation}
The isotropic state $\mathbf Q=\bm 0$ is stable as long as $a>0$ and gets destabilized towards a uniaxial nematic state $S\neq 0$ when $a<0$. Prolate ($S>0$) and oblate ($S<0$) nematic states are coexisting with local stability for $a<0$, and the prolate (oblate) state is energetically favored when $b>0$ ($b<0$). For $a>0$ and $b\neq 0$, a uniaxial nematic state coexists with the isotropic state if $0<a<a_c(b)$ and the isotropic-nematic transition is first order. Only in the case $b=0$, the transition becomes second order.

In two dimensions, the use of the tensor $\mathbf Q_{\mathcal S}$ implies the identity $\mathrm{Tr}[\mathbf Q_{\mathcal S}^3]=0$ if orientations are purely tangential to $\mathcal S$, which imposes a second order transition. In addition, the use of the three-dimensional tensor $\mathbf Q_{\rm tangent}$ from Eq.~\ref{eq-Qt} is consistent with the two-dimensional description since no cubic term appear in $f_{LdG}(\mathbf Q_{\rm tangent})=f_0+f_2\,S_{2D}^2+f_4\,S_{2D}^4$, where $f_0,f_2,f_4$ are coefficients not given explicitly for brevity. Thus, perfect tangential anchoring imposes a second order phase transition whereas unconstrained nematics undergo a first order transition. Applying a soft tangential anchoring condition can therefore control the nature of the transition depending on the anchoring strength.

\subsection{Maier-Saupe free energy}
The use of a statistical definition for the tensors $\mathbf Q$, $\mathbf Q_{\mathcal S}$ makes the eigenvalues bounded. The Landau-de Gennes free energy (Eq.~\ref{eq:fldg}) does not consider such a constraint and is only applicable near the transition region. Since it is easier to define tangential anchoring at the statistical level, by restricting the normal orientation component $u_{\nu}$, we use instead a Maier-Saupe potential~\cite{Maier1958,Fatkullin2005,Ball2010}.

For a distribution of orientations $\rho(\hat{\mathbf u})$, one considers the Maier-Saupe free energy functional
\begin{align}
\mathcal F_{\rm MS}[\rho]=&T\int\mathrm{d}\hat{\mathbf u}\,\rho(\hat{\mathbf u})\log[\rho(\hat{\mathbf u})] \\ \nonumber
&+\frac{1}{2}\int\mathrm{d}\hat{\mathbf u}\int\mathrm{d}\hat{\mathbf v}\,\rho(\hat{\mathbf u})\rho(\hat{\mathbf v})u_{MS}(\hat{\mathbf u},\hat{\mathbf v})
\end{align}
where $T$ is the dimensionless temperature and $u_{MS}$ the Maier-Saupe potential describing the alignment of nematogens, i.e $u_{MS}(\hat{\mathbf u},\hat{\mathbf v})=-\mathbf Q(\hat{\mathbf u}):\mathbf Q(\hat{\mathbf v})$ with $\mathbf Q(\hat{\mathbf u})=(d\,\hat{\mathbf u}\hat{\mathbf u}-\bm 1)/(d-1)$ in $d$ spatial dimensions.

In two dimensions, a second-order phase transition from isotropic to nematic state occurs when $T<1/2$. For a given temperature, the order parameter $S(r)=\mathrm{I}_1(r)/\mathrm{I}_0(r)$ satisfies the implicit equation $S(r)=2rT$ at equilibrium~\cite{Fatkullin2005}, where $r=[0;\infty[$ parametrizes anisotropy of the distribution $\rho(\hat{\mathbf u})$. In three dimensions, only uniaxial states are stable. The isotropic state is stable for $T>2/15\simeq 0.133$, whereas the prolate state is stable for $T<T_c\simeq 0.149$, allowing for a first-order nematic-isotropic transition, Fig.~\ref{fig1}B. For $T<2/15$, the isotropic state transitions into a locally stable oblate state, but the prolate state is the global minimum of the free energy~\cite{Fatkullin2005}.

Now we focus on the effect of soft tangential anchoring. We consider a quadratic anchoring free energy perpendicular to the surface normal $\hat{\bm\nu}$
\begin{equation}
\mathcal F_{\rm anch}[\rho]=W\int\mathrm{d}\hat{\mathbf u}\,\rho(\hat{\mathbf u})(\hat{\mathbf u}\cdot\hat{\bm\nu})^2
\end{equation}
Using spherical coordinates $(\theta,\phi)$ on the unit-sphere with axis $\hat{\bm z}=\hat{\bm\nu}$,
one can show that the total free energy $\mathcal{F}_{\rm tot}=\mathcal{F}_{\rm MS}+\mathcal{F}_{\rm anch}$ is minimized by the following distribution
\begin{equation}
\rho^*(\theta,\phi)=\mathcal{Z}^{-1}\exp[-\tilde{r}(3\cos^2\theta-1)-a\sin^2\theta\cos(2\phi)]
\end{equation}
where $\mathcal{Z}=\int_{0}^{2\pi}\mathrm{d}\phi\int_{0}^{\pi}\mathrm{d}\theta\,\sin\theta\,\mathcal{Z}\rho^*(\theta,\phi)$ by normalization. The distribution is parametrized by two variables, $\tilde{r}=r+W/(3T)$ which describes uniaxial states as a deviation from the isotropic state $\tilde{r}=0$, whereas $a\neq 0$ describes biaxial states.

Studying the local minima of the free energy as a function of the variables $(r,a)$, one finds the phase diagram of Fig.~\ref{fig1}C as a function of reduced temperature $T$ and anchoring coefficient $W$. For $W\neq 0$, the system equilibrates into a planar-isotropic state at high temperature, Fig.~\ref{fig1}D,E. For lower temperatures, the system transitions into a planar-nematic state with non-zero biaxiality, in coexistence with a normal-uniaxial nematic state of higher energy. This normal orientation emerges from a competition between anchoring energy and aligning free energy, such that it is favorable to align uniaxially along the normal direction for small anchoring strength, Fig.~\ref{fig1}C. In addition, coexistence exists between planar-isotropic and planar-nematic states at intermediate temperature and weak anchoring strength, Fig.~\ref{fig1}C,D. The biaxial coefficient $\beta^2=1-6(\mathrm{Tr}[\mathbf Q^3])^2/(\mathbf Q:\mathbf Q)^3$ is shown on Fig.~\ref{fig1}D,E, showing that uniaxiality is a good approximation only far from the isotropic-nematic transition.

A qualitatively similar phase diagram can be obtained using the Landau-de Gennes free energy together with an anchoring energy quadratic in $\mathbf Q$, App.~\ref{app:ldg} and Fig~\ref{figS1}. The coexistence region between planar-isotropic and planar-nematic states can be greatly enhanced with a stiffer anchoring energy, App.~\ref{app:strong-anch} and Fig.~\ref{figS2}. However, this behavior is not conserved in the Maier-Saupe framework when using an anchoring energy $\mathcal{F}_{\rm anch}^{(p)}=W\int\mathrm{d}\hat{\mathbf u}\rho(\hat{\mathbf u})(\hat{\mathbf u}\cdot\hat{\bm\nu})^p$ where $p\leq 6$ is the power controlling the stiffness of the potential, Fig.~\ref{figS3}.

\section{Orientational elastic energies of thin shell nematics}
Next, we focus on orientational elasticity with the free energy density $f_{\rm el}=K|\bm\nabla\mathbf Q|^2/2$ that penalizes spatial gradients of orientation. Here we consider the one-constant approximation on the elastic stiffness $K$ for simplicity. We show using the example of cylindrical geometry that the use of two or three-dimensional tensors, $\mathbf Q_{\mathcal S}$ or $\mathbf Q$, impacts the physics.

\subsection{Orientational elasticity of cylindrical nematic shells}
We consider a nematic shell of cylindrical geometry, with thickness $H$ and mid-plane radius $R_0$. This example is interesting because a cylindrical surface exhibits a coupling of orientational elastic energy $f_{\rm el}$ with extrinsic curvature, as shown in~\cite{Napoli2012a,Napoli2012b,Napoli2013}.

Using cylindrical coordinates $(\rho,\phi,z)$, each value of $\rho=[R_0-H/2;R_0+H/2]$ defines a cylindrical surface $\mathcal S(\rho)$ with outward normal $\hat{\bm\nu}=\hat{\mathbf e}_{\rho}$ and local tangent plane with orthonormal basis $(\hat{e}_{\phi},\hat{e}_z)$. We assume that the shell is sufficiently thin to neglect variations of orientation as a function of $\rho$.

For a tangential director $\hat{\mathbf n}(\phi,z)=n_{\phi}\hat{\mathbf e}_{\phi}+n_z\hat{\mathbf e}_z$, the Frank-Oseen elastic energy~\cite{gennes1993} in the one-constant approximation reads
\begin{equation}\label{eq:fdir}
f_{\rm el}^{\rm dir}=\frac{K_n}{2}|\bm\nabla_{\mathcal S}\hat{\mathbf n}|^2=\frac{K_n}{2}[(\partial_in_z)^2+(\partial_in_{\phi})^2]+\frac{K_n}{2\rho^2}n_{\phi}^2
\end{equation}
where $i=\{\phi,z\}$ and the surface gradient is $\bm\nabla_{\mathcal S}=\frac{1}{\rho}\hat{\mathbf e}_{\phi}\partial_{\phi}+\hat{\mathbf e}_z\partial_z$. The last term on the RHS of Eq.~\ref{eq:fdir} originates from a combination of twist $\sim(\hat{\mathbf n}\cdot\bm\nabla\times\hat{\mathbf n})^2$ and bend $\sim|\hat{\mathbf n}\times\bm\nabla\times\hat{\mathbf n}|^2$ elasticity energies, respectively. It favors the alignment of the director with the longitudinal direction $\hat{\mathbf e}_z$, where the curvature is minimal.

For a purely tangential anchoring, a two-dimensional nematic description corresponds to $\mathbf Q_{\mathcal S}=q_{zz}(\hat{e}_z\hat{e}_z-\hat{e}_{\phi}\hat{e}_{\phi})+q_{z\phi}(\hat{e}_{\phi}\hat{e}_z+\hat{e}_z\hat{e}_{\phi})$ where $q_{\phi\phi}=-q_{zz}$ by construction. Some works~\cite{Pearce2019,Vafa2022,Wang2023} consider the elastic energy to be defined using the covariant derivative $\bm D$ such that
\begin{equation}\label{eq:fcov}
f_{\rm el}^{\rm cov}=\frac{K}{2}|\bm D\mathbf Q_{\mathcal S}|^2=K\left[(\partial_iq_{zz})^2+(\partial_iq_{z\phi})^2\right]
\end{equation}
 for a given $\rho$. To account for extrinsic curvature effects, a coupling term $f_{\rm curv}=K_c[\mathbf Q_{\mathcal S}:(\mathbf c\cdot\mathbf c)]$ is added by hand~\cite{Pearce2019}, where $\mathbf c$ is the curvature tensor. For a cylindrical surface, one finds $f_{\rm curv}=-K_c\,q_{zz}/\rho^2$ which promotes longitudinal alignment ($q_{zz}>0$) for $K_c>0$.

Alternatively, one also finds in the literature~\cite{Virga2011,Virga2015,Salbreux2022,Mesarec2023} the use of the surface derivative $\bm\nabla_{\mathcal S}$ such that the elastic energy becomes
\begin{equation}\label{eq:fsurf}
f_{\rm el}^{\rm surf}=\frac{K}{2}|\bm\nabla_{\mathcal S}\mathbf Q_{\mathcal S}|^2=f_{\rm el}^{\rm cov}+\frac{K}{\rho^2}(q_{zz}^2+q_{z\phi}^2)
\end{equation}
where the last term originates from the basis vector gradient $\partial_{\phi}\hat{\mathbf e}_{\phi}=-\hat{\mathbf e}_{\rho}$.

Finally a three-dimensional description with tangential anchoring ($u_{\nu}=0$) gives $\mathbf Q_{\rm tangent}=\frac{1}{4}(1+3q_{zz})\hat{e}_z\hat{e}_z+\frac{1}{4}(1-3q_{zz})\hat{e}_{\phi}\hat{e}_{\phi}+\frac{3}{4}q_{z\phi}(\hat{e}_{\phi}\hat{e}_z+\hat{e}_z\hat{e}_{\phi})-\frac{1}{2}\hat{\mathbf e}_{\rho}\hat{\mathbf e}_{\rho}$ using Eq.~\ref{eq-Qt}. The elastic energy becomes
\begin{equation}\label{eq:f3d}
f_{\rm el}^{\rm 3D}=\frac{K}{2}|\bm\nabla_{\mathcal S}\mathbf Q_{\rm tangent}|^2=\frac{9}{16}f_{\rm el}^{\rm cov}+\frac{9K}{16\rho^2}[(q_{zz}-1)^2+q_{z\phi}^2]
\end{equation}
In all cases the total elastic free energy is $F_{\rm el}=\int\mathrm{d}z\mathrm{d}\phi\mathrm{d}\rho\,\rho f_{\rm el}$.

The four levels of description from Eqs.~\ref{eq:fdir}-\ref{eq:f3d} contain uniform terms that represent a coupling to extrinsic curvature. The approach of Eq.~\ref{eq:fcov} considers that thin nematic shells and nematic surfaces have distinct physics, such that phenomenological terms can be added with dependence on the details of the surface anchoring properties~\cite{Biscari2006}. Eq.~\ref{eq:fdir} represents the thin shell limit of a nematic liquid crystal in the limit of uniform nematic order. Both Eqs.~\ref{eq:fsurf},\ref{eq:f3d} are candidates for a $\mathbf Q$-tensor theory of nematic thin shells. Logically, this theory should admit Eq.~\ref{eq:fdir} as a limit case when the nematic order is uniform.

One can easily show that $q_{zz}^2+q_{z\phi}^2=S_{2D}^2$ whereas $(q_{zz}-1)^2+q_{z\phi}^2=(1-S_{2D})^2+4S_{2D}n_{\phi}^2$. Thus, Eq.~\ref{eq:f3d} reproduces the expected coupling to extrinsic curvature from the director theory of \cite{Napoli2012a} with the equivalence $K_n=9KS_{2D}/2$. In addition, it contains an ordering term $\sim K(1-S_{2D})^2/\rho^2$ already acknowledged in~\cite{Napoli2012b} that renormalizes $f_{LdG}$ or $f_{MS}$. This effect directly originates from the tangential restriction of the three-dimensional tensor, Eq.~\ref{eq-Qt}. It comes from an energy penalty to align along $\hat{\mathbf e}_{\phi}$ such that nematogens get ordered along $\hat{\mathbf e}_z$. As noted in~\cite{Napoli2012b}, this ordering term depends only on curvature and prevents purely isotropic states $S_{2D}=0$ for surfaces with vanishing Euler-Poincare characteristic like a cylinder. Note again that Eq.~\ref{eq-Qt} uses the statistical expression of $\mathbf Q$ where eigenvalues are bounded by unity, whereas the Laudau-de Gennes free energy does not impose such a constraint.

In contrast, Eq.~\ref{eq:fsurf} does not contain the expected coupling to extrinsic curvature aligning the director along $\hat{\mathbf e}_z$, but prevents nematic order for highly curved cylinders, $\sim KS_{2D}^2/\rho^2$. This distinction was already acknowledged in~\cite{Napoli2012b,Napoli2013} in the limit of perfect tangential anchoring, without physical explanation for this discrepancy.
We discussed above that the limit from weak to strong tangential anchoring needs a continuity of descriptions with the three-dimensional tensor $\mathbf Q$. Thus, one concludes that the uniform terms in Eq.~\ref{eq:fsurf} originate from a mathematical artefact of the two-dimensional tensor $\mathbf Q_{\mathcal S}$ over a curved surface, without physical meaning. One can add by hand a term coupling nematic order to the invariants of curvature, but this phenomenological approach allows a priori positive or negative coupling constants. In opposition, the three-dimensional version (Eq.~\ref{eq:f3d}) predicts that curvature promotes nematic ordering for a cylinder and is compatible with the limit case of Eq.~\ref{eq:fdir}.
To summarize, we showed using the example of a cylindrical surface that a three-dimensional $\mathbf Q$-tensor is required to describe thin nematic shells.


\subsection{Experimental test for two- or three-dimensional descriptions of nematic shells}
For active systems with nematic symmetry, the mechanics of the shell is often coupled to orientational order~\cite{Sanchez2012,Marchetti2013,Keber2014,AlIzzi2023,Dessalles2025}. Then, the difference between two- and three- dimensional descriptions translates into experimentally measurable effects.

The active stress $\bm\sigma^{\rm (a)}$ is then proportional to the nematic tensor, with $\bm\sigma^{\rm (a)}=\alpha\mathbf Q$ for three-dimensional nematic activity or $\bm\sigma^{\rm (a)}=\alpha\mathbf Q_{\mathcal S}$ for activity restricted to the tangent plane. Under a thin film approximation (see Section V for details), the stress divergence defines an active force density $\bm f^{\rm (a)}=\bm\nabla\cdot\bm\sigma^{\rm (a)}$ applied on the surroundings by the material.

Again, we use cylindrical geometry as an illustrative example where $\mathbf Q_{\mathcal S}=q_{zz}(\hat{\mathbf e}_z\hat{\mathbf e}_z-\hat{\mathbf e}_{\phi}\hat{\mathbf e}_{\phi})+q_{z\phi}(\hat{\mathbf e}_z\hat{\mathbf e}_{\phi}+\hat{\mathbf e}_{\phi}\hat{\mathbf e}_z)$. We consider the nematic shell to be in contact with an elastic substrate on its inner surface. The coupling to extrinsic curvature leads to the uniform terms $\bm\nabla\cdot\mathbf Q|_{\rm uniform}=-3(1-q_{zz})/4R\,\hat{\mathbf e}_{\rho}$ and $\bm\nabla\cdot\mathbf Q_{\mathcal S}|_{\rm uniform}=q_{zz}/R\,\hat{\mathbf e}_{\rho}$.

Thus, a contractile material with $\alpha>0$ and longitudinal orientation $0<q_{zz}\leq 1$ produces an outward radial force on the substrate in the two-dimensional description, but an inward radial force in the three-dimensional one. In contrast, an azimuthal orientation with $-1\leq q_{zz}<0$ produces inward radial forces on the substrate in both cases, which is expected for contractile rings. The difference originates from a non-zero active radial stress in the three-dimensional description $\sigma_{\rho\rho}^{\rm (a)}=-\alpha/2\,\hat{\mathbf e}_{\rho}\hat{\mathbf e}_{\rho}$. Indeed, it was recently shown theoretically in Ref.~\cite{Nejad2025} that the combination of in-plane active contraction and out-of-plane active pressure can lead to a thickness instability of a suspended cylindrical film with longitudinal nematic alignment.

Importantly, only the limit of stiff elastic substrates allows to eliminate radial force balance. Otherwise, the normal forces applied by the nematic shell must be taken into account.
This effect has direct experimental consequences for active nematic shells deposited on soft cylindrical gels~\cite{Dessalles2025}. In this reference, the authors consider an endothelial monolayer adhered to a deformable cylindrical shell, and induce evolution of its radius through controlled luminal pressure changes. They observe a transition from longitudinal to transverse cell orientation upon application of luminal pressure. Their theory uses the two-dimensional tensor $\mathbf Q_{\rm S}$ from Eq.~\ref{eq:q2d}, but assumes a uniaxial active stress tensor $\bm\sigma^{\rm a}=\alpha(\mathbf Q_{\rm S}+\bm 1_{\mathcal S})/2=\alpha\langle\mathbf u_{\mathcal S}\mathbf u_{\mathcal S}\rangle$ with $\alpha>0$. This implies an active force density $\bm\nabla\cdot\bm\sigma^{(a)}|_{\rm uniform}=-\alpha(1-q_{zz})/2R\,\hat{\mathbf e}_{\rho}$ pushing the substrate shell inward for any nematic orientation $-1\leq q_{zz}\leq 1$. Up to a numerical factor, this corresponds exactly to the radial active force from a three-dimensional nematic tensor, $\bm\nabla\cdot\bm\sigma^{(a)}|_{\rm uniform}=-3\alpha(1-q_{zz})/4R\,\hat{\mathbf e}_{\rho}$. Thus, the good agreement between experiments and theory in Ref.~\cite{Dessalles2025} provides an indirect experimental evidence for the relevance of a three-dimensional tensor description of thin nematic shells.

\section{Q-tensor theory for an adhered nematic shell}
Now we construct a mechanical theory of three-dimensional nematohydrodynamics under the thin film approximation. For simplicity, we restrict our study to incompressible materials with purely viscous rheology and active nematic stress. We consider the effects of weak anchoring and boundary anisotropy in presence of a solid substrate on one side and a free interface on the other side. This has direct relevance for many experimental systems of living matter~\cite{Sanchez2012,Keber2014,Sano2017,Ravichandran2025,Dessalles2025}.

\subsection{Nematohydrodynamic equations}
\begin{figure}
\centering
\includegraphics[width=1.\linewidth]{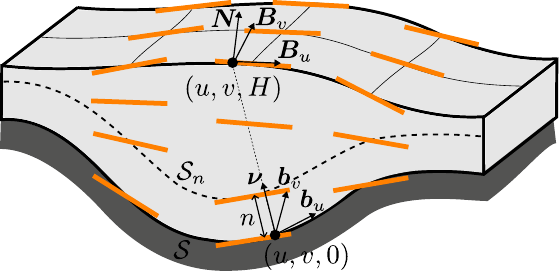}
\caption{Schematics of a thin nematic shell (light gray) adhered to a substrate (dark gray). At the surface of the substrate $\mathcal S$, surface coordinates $(u,v)$ with tangent vectors $(\mathbf b_u,\mathbf b_v)$ and normal vector $\hat{\bm\nu}$ are used. A level set of $\mathcal S$ is used with normal coordinate $n$, defining the surface $\mathcal S_{n}$ with the substrate at $n=0$. Each material point is described with the coordinates $(u,v,n)$. The free surface of the material is at $n=H(u,v)$ with interfacial tangent vectors $(\bm B_u,\bm B_v)$ and interfacial normal vector $\bm N$. Orange: nematic orientation at various spatial points.
}
\label{fig2}
\end{figure}

In this section, we consider a reference surface $\mathcal S$ parametrized by coordinates $(u,v)$, corresponding to the substrate on top of which the nematic material resides. We use the normal coordinate $n$ along the outward normal $\hat{\bm\nu}$ to construct level sets with surfaces $\mathcal S_{n}$, such that $\bm r=\bm r_{\mathcal S}+n\hat{\bm\nu}$ with the substrate localized at $n=0$, Fig.~\ref{fig2}. One can construct an orthogonal basis $\{\mathbf b_u(n),\mathbf b_v(n),\hat{\bm\nu}\}$ at any level $n$ using the directions defined by the principal curvatures $c_i$, where $i=\{u,v\}$. One finds the tangent vectors $\mathbf b_i(n)=(1+n c_i)\mathbf b_i(0)$ and gets the metric components $g_{ij}(n)=\mathbf b_i(n)\cdot\mathbf b_j(n)=\mathrm{Diag}[E_{n},G_{n}]$, App.~\ref{app:derivatives}. All tensorial fields are expressed in terms of the orthonormal basis $\{\hat{\mathbf e}_u,\hat{\mathbf e}_v,\hat{\bm\nu}\}$ with $\hat{\mathbf e}_i=\mathbf b_i/|\mathbf b_i|$. At the free interface, the surface is parametrized by $n=H(u,v,t)$ where $H$ is the thickness, and defines interfacial tangent vectors $\bm B_i(u,v)=[1+H\,c_i]\mathbf b_i(0)+\partial_iH\hat{\bm\nu}$, Fig.~\ref{fig2}. The free surface normal follows as $\bm N=(\bm B_u\times\bm B_v)/|\bm B_u\times\bm B_v|$ and defines the interface curvature components $C_{ij}=\bm B_i\cdot\partial_j\bm N$, App.~\ref{app:free-surface}. We use the convention where the curvatures of a locally spherical surface are positive.

In three dimensions, the mechanics of an incompressible material emerges from a stress tensor
\begin{align}
\bm\sigma=-P\bm 1+2\eta\tilde{\mathbf u}+\alpha\mathbf Q
\end{align}
where $P$ is the pressure, $\eta$ the shear viscosity, $\tilde{\mathbf u}=\frac{1}{2}(\bm\nabla\mathbf v+\bm\nabla\mathbf v^T)$ the traceless strain rate with the velocity $\mathbf v$, and $\alpha$ the active stress coefficient. We assumed here that the active stress is sufficiently large to neglect contributions from Ericksen and reactive stresses. Bulk force balance then reads
\begin{align}
\bm\nabla\cdot\bm\sigma=-\bm\nabla P+\eta\Delta\mathbf v+\alpha\bm\nabla\cdot\mathbf Q=\bm 0
\end{align}
with incompressibility $\bm\nabla\cdot\mathbf v=0$. All fields are function of the curvilinear coordinates $(u,v,n)$. In addition, the dynamics of the nematic tensor $\mathbf Q$ is controlled by
\begin{align}
(\partial_t+\mathbf v\cdot\bm\nabla)\mathbf Q+\bm\omega\cdot\mathbf Q-\mathbf Q\cdot\bm\omega=\frac{\bm H}{\Gamma}+\frac{3\lambda}{2}\tilde{\mathbf u}
\end{align}
where $\bm\omega=\frac{1}{2}(\bm\nabla\mathbf v-\bm\nabla\mathbf v^T)$ is the vorticity tensor, $\bm H$ the molecular field, $\Gamma$ the rotational viscosity and $\lambda$ the flow-alignment coefficient. The molecular field is $\bm H=-\delta\mathcal F/\delta\mathbf Q=K\Delta\mathbf Q-\partial_{\mathbf Q}f_B(\mathbf Q)$. The bulk part of the free energy density $f_B$ can be of Landau-de Gennes or Maier-Saupe format, with $\mathcal F=\int\mathrm{d}V[f_B(\mathbf Q)+f_{\rm el}(\bm\nabla\mathbf Q)]$. Over the surface $\mathcal S$, we add an anchoring free energy $\mathcal F_{\rm anch}=\int_{\mathcal S}\mathrm{d}A\,f_{\rm anch}(\mathbf Q)$ at the substrate interface $n=0$. A first variation of the total free energy $\mathcal F+\mathcal F_{\rm anch}$ gives the natural boundary condition $K(\hat{\bm\nu}\cdot\bm\nabla)\mathbf Q-\partial_{\mathbf Q}f_{\rm anch}=0$. For the quadratic expression used in the appendix, Eq.~\ref{eq:f-anch}, one finds
\begin{align}
\partial_{\mathbf Q}f_{\rm anch}=\frac{w}{6}[(1+2Q_{\nu\nu})(2\hat{\bm\nu}\hat{\bm\nu}-\hat{\mathbf e}_i\hat{\mathbf e}_i)+3Q_{\nu i}(\hat{\mathbf e}_i\hat{\bm\nu}+\hat{\bm\nu}\hat{\mathbf e}_i)]
\end{align}
with anchoring coefficient $w$ and summation over repeated indices $i$. In the limit of perfect tangential anchoring $w\rightarrow\infty$, one obtains $Q_{u\nu}=Q_{v\nu}=Q_{\nu\nu}+1/2=0$.

Boundary conditions at the free interface $n=H$ are an absence of shear stress $\bm N\cdot\bm\sigma\cdot\bm B_i=0$, Laplace pressure $\bm N\cdot\bm\sigma\cdot\bm N=-\gamma(C_1+C_2)$ from surface tension $\gamma$, and nematic free anchoring $K(\bm N\cdot\bm\nabla)\mathbf Q=\bm 0$. Surface tension should be considered anisotropic in all generality but we keep it isotropic for simplicity. A kinematic condition of the interface motion is derived from $(\partial_t+\mathbf v\cdot\bm\nabla)F=0$ with the implicit surface function $F(u,v,n,t)=H(u,v,t)-n$. Finally, boundary conditions on the substrate at $n=0$ are normal impenetrability $\mathbf v\cdot\hat{\bm\nu}=0$, viscous tangential friction $\hat{\bm\nu}\cdot\bm\sigma\cdot\hat{\mathbf e}_i=\xi_s\mathbf v\cdot\hat{\mathbf e}_i$ and weak tangential anchoring $K(\hat{\bm\nu}\cdot\bm\nabla)\mathbf Q=\partial_{\mathbf Q}f_{\rm anch}$.

\subsection{Thin film expansion}
We can simplify the previous set of equations by assuming the shell thickness $H$ to be much smaller than the other dimensions of the system characterized by $L$. The thin film parameter $\epsilon\sim H/L$, and principal curvatures $c_i\sim\mathcal{O}(1)$ are small enough for all material points to be uniquely defined by the coordinate set $(u,v,n)$. This implies $E_{n}=E_0\sim\mathcal{O}(1)$, $G_{n}=G_0=\mathcal{O}(1)$ at lowest order.

They are three physical length scales associated to the normal direction, the hydrodynamic length $L_h=\eta/\xi_s$, the anchoring length $L_w=K/w$ and the active length $L_a=\gamma/|\alpha|$. A thin film expansion is appropriate if all the length scales of the problem are larger than $H$. Importantly, note that the usual limit of strong anchoring ($w\rightarrow\infty$) implies $L_w\leq H$ and enters in contradiction with a thin film approximation, inducing strong coupling between normal and tangential equations. To obtain a decoupling of those directions in the equations, we assume the scaling assumptions $H/L_h,H/L_w\sim\mathcal{O}(\epsilon^2)$ and $H/L_a\sim\mathcal{O}(\epsilon)$, or $\xi_s\sim\epsilon\,\eta/L$, $\gamma\sim|\alpha|L$ and $w\sim\epsilon\,K/L$. A scaling $H/L_a\sim\mathcal{O}(\epsilon^2)$ is not compatible with boundary conditions for an incompressible material with curvatures $c_i\sim\mathcal{O}(1)$. The case of a flat substrate $c_u=c_v=0$ is discussed in App.~\ref{app:lsa-flat}.

We expand all the fields $\phi=\{\mathbf v,\mathbf Q,P\}$ as $\phi(u,v,n,t)=\phi^{(0)}(u,v,n,t)+\epsilon\,\phi^{(1)}(u,v,n,t)+\mathcal{O}(\epsilon^2)$. First, incompressibility implies $v_{\nu}^{(0)}=0$ at lowest order by impermeability with the substrate. At next order, one gets the incompressibility equation
\begin{equation}\label{eq:incomp0}
\frac{\partial_uv_u^{(0)}-a_{uu}^{(0)}v_v^{(0)}}{\sqrt{E_0}}+\frac{\partial_vv_v^{(0)}-a_{vv}^{(0)}v_u^{(0)}}{\sqrt{G_0}}+\partial_{n}v_{\nu}^{(1)}=0
\end{equation}
The geometric prefactors $a_{uu},a_{vv}$ are detailed in Eq.~\ref{eq:aii}. In parallel, force balance equations reduce to $\eta\partial_{n}^2v_i^{(0)}=0$ in the tangent plane and $-\partial_{n}P^{(0)}+\eta\partial_{n}^2v_{\nu}^{(1)}+\alpha\partial_{n}Q_{\nu\nu}^{(0)}=0$ along the normal direction, at lowest order. The corresponding stress boundary conditions give $\partial_{n}v_i^{(0)}=0$ at $n=0,H$, see App.~\ref{app:bcs} for details. This implies that tangential velocity components $v_i^{(0)}$ are independent from $n$, and one obtains $v_{\nu}^{(1)}(n)=n\,\partial_{n}v_{\nu}^{(1)}$ with the prefactor obtained from Eq.~\ref{eq:incomp0}.

The dynamic equations of $\mathbf Q$ together with boundary conditions give $\partial_{n}\mathbf Q^{(0)}=\partial_{n}\mathbf Q^{(1)}=\bm 0$ at the two lowest orders, and normal force balance is reduced to $\partial_{n}P^{(0)}=0$.
Stress boundary conditions give
\begin{equation}\label{eq:pressure}
P^{(0)}=2\eta\partial_{n}v_{\nu}^{(1)}+\alpha Q_{\nu\nu}^{(0)}+\gamma(C_1^{(0)}+C_2^{(0)}).
\end{equation}
and $\eta\partial_{n}v_i^{(1)}=\eta c_iv_i^{(0)}-\alpha Q_{\nu i}^{(0)}$.

The first non-trivial tangential equations appear at order $\mathcal{O}(1)$
\begin{align}
&\eta\,\hat{\mathbf e}_i\cdot\Delta\mathbf v^{(0)}
+\eta(c_u+c_v)\partial_{n}v_i^{(1)}+\eta\partial_{n}^2v_i^{(2)} \\ \nonumber
&=\nabla_{\mathcal S,i}P^{(0)}-\alpha(\bm\nabla\cdot\mathbf Q^{(0)})_i \\
\partial_t\mathbf Q^{(0)}&+\mathbf v^{(0)}\cdot[\bm\nabla\mathbf Q]^{(0)}+\bm\omega^{(0)}\cdot\mathbf Q^{(0)}-\mathbf Q^{(0)}\cdot\bm\omega^{(0)} \\ \nonumber
&=\frac{K}{\Gamma}(\Delta\mathbf Q^{(0)}+\partial_{n}^2\mathbf Q^{(2)})-\frac{\partial_{\mathbf Q}f_B^{(0)}}{\Gamma}+\frac{3\lambda}{2}\tilde{\mathbf u}^{(0)}
\end{align}
where $\nabla_{\mathcal S,i}=\hat{\mathbf e}_i\cdot\bm\nabla_{\mathcal S}=|\mathbf b_i(0)|^{-1}\partial_i$ is the tangential gradient at $n=0$.
Expressions of pressure gradient, velocity Laplacian, vorticity, strain rate nematic gradient, nematic Laplacian are given in App.~\ref{app:derivatives0}. From the previous relations, the tangential bulk equations imply $\partial_{n}^3v_i^{(2)}=0$ and $\partial_{n}^3\mathbf Q^{(2)}=\bm 0$.
At next order, nematic boundary conditions are $K\partial_{n}\mathbf Q^{(2)}(0)=\partial_{\mathbf Q}f_{\rm anch}^{(1)}$ at $n=0$ and $K\partial_{n}\mathbf Q^{(2)}(H)=K\nabla_{\mathcal S,j}H\,\hat{\mathbf e}_j\cdot[\bm\nabla\mathbf Q]^{(0)}$ at $n=H$. Using the identity $H\,\partial_{n}^2\mathbf Q^{(2)}=\partial_n\mathbf Q^{(2)}(H)-\partial_n\mathbf Q^{(2)}(0)$,
one finally obtains effective tangential dynamics of the three-dimensional nematic tensor
\begin{align}\label{eq:qtangent}
&\partial_t\mathbf Q+\bm\omega\cdot\mathbf Q-\mathbf Q\cdot\bm\omega
+\left[\mathbf v-\frac{K\bm\nabla_{\mathcal S}H}{\Gamma H}\right]\cdot\bm\nabla\mathbf Q \\ \nonumber
&=\frac{K}{\Gamma}\Delta\mathbf Q-\frac{\partial_{\mathbf Q}f_B}{\Gamma}+\frac{3\lambda}{2}\,\tilde{\mathbf u} \\ \nonumber
&-\frac{w}{6\Gamma H}[(1+2Q_{\nu\nu})(2\hat{\bm\nu}\hat{\bm\nu}-\hat{\mathbf e}_i\hat{\mathbf e}_i)+3Q_{\nu i}(\hat{\mathbf e}_i\hat{\bm\nu}+\hat{\bm\nu}\hat{\mathbf e}_i)]
\end{align}
where the superscript ${(0)}$ has been dropped for clarity.

For a passive material, the last term on the RHS is an anchoring torque which tends to relax the normal components of the nematic tensor $\mathbf Q\cdot\hat{\bm\nu}$ towards the tangential state $Q_{\nu i}=Q_{\nu\nu}+1/2=0$, with a characteristic time scale $\tau_w=\Gamma H/w$. The anchoring torque competes with the uniform part of the molecular field $-\partial_{\mathbf Q}f_B$. We write $\partial_{\mathbf Q}f_B=\chi\,c(\mathbf Q)\mathbf Q$ where $c$ is a non-linear function of $\mathbf Q$ and $\chi$ an ordering coefficient. Hence, the tangential anchoring limit is valid when $w\gg\chi H$, true in particular for very thin nematic shells. In contrast, in presence of a non-zero active stress, the flow-alignment term can generate a drift term and prevent relaxation towards the tangential nematic state. Indeed, one finds $\partial_tQ_{\nu i}\sim-\lambda\alpha Q_{\nu i}/\eta$ at high activity coefficient $\alpha$ such that $\tilde{\mathbf u}\sim-\alpha\mathbf Q/\eta$. Contractile activity $\alpha>0$ reinforces the tangential relaxation, but an instability is expected for extensile activity when $\alpha H\lesssim -w\eta/(\lambda\Gamma)$. Thus, an assumption of perfect tangential orientation is limited to some subregion of the parameter space.

In addition, from the identities $H\,\partial_{n}^2v_i^{(2)}=\partial_nv_i^{(2)}(H)-\partial_nv_i^{(2)}(0)$ and $H\partial_nv_i^{(1)}=v_i^{(1)}(H)-v_i^{(1)}(0)$ with shear-stress boundary conditions, App.~\ref{app:bcs}, one finally obtains effective tangential force balance
\begin{align}\label{eq:ftangent}
&\eta H[(\Delta\mathbf v)_i+c_i^2v_i+3\nabla_{\mathcal S,i}(\bm\nabla_{\mathcal S}\cdot\mathbf v_{\mathcal S})] \\ \nonumber
&+\alpha H[(\bm\nabla_{\mathcal S}\cdot\mathbf Q)_i-(c_u+c_i+c_v)Q_{\nu i}-\nabla_{\mathcal S,i}Q_{\nu\nu}] \\ \nonumber
&+[\eta(2\tilde{u}_{ij}+3\bm\nabla_{\mathcal S}\cdot\mathbf v_{\mathcal S}\,\delta_{ij})+\alpha(Q_{ij}-Q_{\nu\nu}\delta_{ij})]\nabla_{\mathcal S,j}H \\ \nonumber
&=[\xi_s-\eta H\,c_g]v_i+\gamma H\nabla_{\mathcal S,i}(c_u+c_v)
\end{align}
The superscript ${(0)}$ has been dropped for clarity
, $\mathbf v_{\mathcal S}=v_i\hat{\mathbf e}_i$ is the tangential projection of the velocity field, $c_g=c_uc_v$ is the Gaussian curvature, and the pressure has been replaced by its expression, Eq.~\ref{eq:pressure}. One obtains a tangential hydrodynamic length $L_h=\sqrt{\eta H/\xi_s}$, an effective planar bulk viscosity $\eta_b H=3\eta H$ and surface activity coefficient $\alpha_h=\alpha H$. The friction is locally renormalized as $\xi_s\mapsto\xi_s-\eta H c_g$, and illustrates the idea that tangential velocity at a substrate point is no longer tangential in the neighborhood because of curvature. Note that surface tension does not impact height modulations at least order for a curved substrate.

In the limit of tangential nematic anchoring $Q_{\nu i}=1/2+Q_{\nu\nu}=0$, one finds free surface gradients $\bm\nabla_{\mathcal S}H$ to drive anisotropic flows with $\xi_sv_i\sim\alpha[Q_{ij}+1/2\delta_{ij}]\nabla_{\mathcal S,j}H$, suggesting a shape instability at high activity. As shown in App.~\ref{app:free-surface}, the height function follows the dynamic equation
\begin{equation}\label{eq:dtH}
\partial_tH+\mathbf v_{\mathcal S}\cdot\bm\nabla_{\mathcal S}H=-H(\bm\nabla_{\mathcal S}\cdot\mathbf v_{\mathcal S})
\end{equation}
at first order.
We then perform linear stability analysis around the state $H=H_0$, $\mathbf v=\bm 0$, $\mathbf Q=\mathbf Q_{\rm tangent}$ for a flat substrate. We find a thickness shape instability to be triggered by contractile active coefficient $\alpha>0$ along the director orientation, with a characteristic length scale $\tilde{\lambda}=\sqrt{\gamma H_0/\alpha}$, see App.~\ref{app:lsa-flat}. Interestingly, it also triggers the instability of the out-of-plane nematic component $Q_{\nu\nu}$ through flow-alignment since $\tilde{u}_{\nu\nu}=-\bm\nabla_{\mathcal S}\cdot\mathbf v_{\mathcal S}$. This makes a full three-dimensional description indispensable as long as $w\lesssim\Gamma|\alpha|H/\eta$.

Even in the limit of strong tangential anchoring $w\rightarrow\infty$ at the substrate imposing $Q_{\nu\nu}+1/2=Q_{\nu i}=0$ for thin shells, the presence of a free surface at $n=H$ induces non-trivial effects. Assuming that there exists a parametric regime where the thin film condition $L_w>H$ remains valid, the factor $w(1+2Q_{\nu\nu})$ is transformed into a Lagrange multiplier $\Lambda$ such that
\begin{align}\label{eq:Lambda}
\Lambda&=3KH[c_u^2(1+2Q_{uu})+c_v^2(1+2Q_{vv})] \\ \nonumber
&+\frac{3\chi H}{2}\,c(\mathbf Q)-\frac{9H\Gamma\lambda}{2}(\bm\nabla_{\mathcal S}\cdot\mathbf v_{\mathcal S})
\end{align}
The Lagrange multiplier $\Lambda$ replaces the factor $w(1+2Q_{\nu\nu})$ in the dynamics of diagonal components $Q_{ii}\hat{\mathbf e}_i\hat{\mathbf e}_i$, Eq.~\ref{eq:qtangent}, for $\mathbf Q$ to remain traceless. This approach can be formalized through the introduction of Lagrange multipliers in $f_{\rm anch}$, see App.~\ref{app:strong-anch}. The effect of the Lagrange multiplier $\Lambda$ is to introduce out-of-plane coupling in the nematic dynamics. It contributes when substrate curvature is non-zero by orientation mismatch at boundaries, when the bulk free energy factor $c(\mathbf Q)$ is not compatible with tangential anchoring, or when non-zero flow divergence deforms the free interface, see Eq.~\ref{eq:dtH}. Again, a shape instability fed by active stresses could emerge for $\bm\nabla_{\mathcal S}\cdot\mathbf v_{\mathcal S}<0$ and contribute to nematic evolution through the Lagrange multiplier $\Lambda$.

\subsection{Active nematic shell on a sinusoidal substrate}
As an application of the previous set of thin film equations, we study the spontaneous flow transition driven by active nematic stress for a sinusoidal substrate~\cite{Bell2022}. Curved substrates can be realized experimentally for cell monolayers~\cite{Luciano2021,Harmand2022}, and show mechanosensitive mechanisms adapting cell orientation to principal curvatures. These results motivated theoretical work where nematic orientation is coupled to curvature in a minimal way~\cite{Bell2022}. 

Within our framework, this coupling would be equivalent to the free energy density $f_{\rm curv}=\frac{1}{2}K_c(\mathbf Q\cdot\mathbf c):\mathbf Q$ on the substrate surface $\mathcal S$ with curvature tensor $\mathbf c$. It yields a boundary torque proportional to $\partial_{\mathbf Q}f_{\rm curv}=K_c\,[\frac{1}{2}(\mathbf Q\cdot\mathbf c+\mathbf c\cdot\mathbf Q)-\frac{1}{3}(\mathbf Q:\mathbf c)\bm 1]$.
Then, the thin film expansion of Sec.~V,B is modified through the substrate boundary condition $K\bm\partial_n\mathbf Q^{(2)}(0)=\partial_{\mathbf Q}f_{\rm anch}+\partial_{\mathbf Q}f_{\rm curv}$. As before, we assume $K_c\sim\epsilon\,K$ for the new length scale $L_c=K/K_c$ to satisfy the thin film condition $H/L_c\sim\mathcal{O}(\epsilon^2)$. The new coupling enters into nematic dynamics by adding the term $-\partial_{\mathbf Q}f_{\rm curv}/(\Gamma H)$ on the RHS of Eq.~\ref{eq:qtangent}.
For finite tangential anchoring strength $w$, the effect of $f_{\rm curv}$ is to destabilize the tangential state when $K_c c_i<0$.

We now focus on a sinusoidal substrate localized at $z=c_0L^2\cos(2\pi u/L)$ with arc-length $u=[0,L]$, characteristic substrate curvature $c_0$ and invariance along $y=v$. We consider strong tangential anchoring $w\rightarrow\infty$ and deep nematic phase $\chi\rightarrow\infty$, with an initial orientation state $\mathbf{Q}_0=\hat{\mathbf e}_v\hat{\mathbf e}_v-\frac{1}{2}(\hat{\mathbf e}_u\hat{\mathbf e}_u+\hat{\bm\nu}\hat{\bm\nu})$ in the longitudinal direction $\hat{\mathbf e}_v$. Performing linear stability analysis around this static state with strong anchoring at edges $u=0,L$, we find a shear flow instability triggered by extensile active stress $\alpha(\lambda-1)<0$ and $K_c c_0>0$, App.~\ref{app:lsa-sin}. On the opposite, sinusoidal curvature also stabilizes longitudinal alignment through orientational elasticity $K$ (Sec.~IV). The shell thickness also remains constant in that case. This is coherent with the results from Ref.~\cite{Bell2022}, and it shows that the tangential state is further stabilized when $K_c c_0>0$.

For a transverse mode with an initial orientation state $\mathbf{Q}_0=\hat{\mathbf e}_u\hat{\mathbf e}_u-\frac{1}{2}(\hat{\mathbf e}_v\hat{\mathbf e}_v+\hat{\bm\nu}\hat{\bm\nu})$, linear stability analysis gives the standard shear flow instability for $\alpha(\lambda+1)<0$ and $K_cc_0>0$ as before. However, the transverse velocity $v_u$ now couples with the thickness $H=H_0+\delta H$ and drives an instability for contractile activity $\alpha>0$, App.~\ref{app:lsa-sin}. At linear level, there is no coupling between substrate curvature and thickness growth. Yet, a modulation of shell thickness depending on curvature is observed in experiments~\cite{Luciano2021,Harmand2022}, and proposed to arise from asymmetric apico-basal activity. For a simple active layer, it would be interesting to study the impact of non-linearities on the thickness-curvature coupling.

\section{Discussion}
For theoretical descriptions of quasi-two dimensional liquid crystals, it is customary to restrict the orientation of nematogens to the tangent plane and use the tensor $\mathbf Q_{\mathcal S}$. A priori, this restriction implicitly assumes the presence of a strong physical effect to prevent any out-of-plane deviation of orientations. Then, the isotropic-nematic transition must be of second order in two dimensions, whereas it can be of first order in three dimensions.
By relaxing this hypothesis and considering weak tangential anchoring, we allow for a connection between the two descriptions and one must use the three-dimensional tensor $\mathbf Q$ . We find that a coexistence region exists between planar-isotropic and planar-nematic states, as expected from a first order phase transition. This result is valid for any shell thickness as long as nematic gradients do not enter into play, in particular for infinitely thin layers, surfaces. Although the coexistence region is narrow in parameter space, active systems like cytoskeletal filament suspensions might amplify fluctuations of orientation, weaken tangential anchoring and promote such two-dimensional first-order phase transition.

Even if perfect tangential anchoring is a good approximation and if the shell is infinitely thin, we showed for passive systems and systems with active nematic stress that the use of the tensor $\mathbf Q_{\mathcal S}$ introduces physical artefacts. Indeed when a nematic shell is in contact with a deformable substrate, out-of-planes forces become experimentally relevant and the coupling to geometry produces physically distinct forces when $\mathbf Q_{\mathcal S}$ or $\mathbf Q$ are used. Either in terms of nematic-curvature coupling or in terms of out-of-plane active forces, our results show that a theoretical description using the tensor $\mathbf Q_{\mathcal S}$ is only relevant for limited cases.

Finally, we derived a theory of nematohydrodynamics for thin curved shells in contact with a substrate, which has many applications in biological physics, for the cytoskeletal cortex of cells or tissues of elongated cells. Again, we assumed weak tangential anchoring of nematogens. Remarkably, we found that the presence of active stresses can destabilize tangential orientation for both contractile or extensile active stress, through different mechanisms. These results set a lower bound on the magnitude of the tangential anchoring coefficient $w$ for those active effects to be negligible. All together, our results highlight the need to consider full three-dimensional descriptions of nematic orientation for curved nematic shells. We expect those considerations to be essential for studying important morphological processes of living organisms and deepen our understanding of nematic materials.

\begin{acknowledgments}
I thank K. Kruse for insightful discussions and careful reading of the manuscript.
\end{acknowledgments}

\appendix
\setcounter{figure}{0}
\renewcommand{\thefigure}{A\arabic{figure}}

\section{Isotropic-to-nematic transition with the Landau-de Gennes energy}\label{app:ldg}
We start from the Landau-de Gennes free energy density
\begin{equation}
f_{LdG}(\mathbf Q)=\frac{a}{2}\mathrm{Tr}[\mathbf Q^2]-\frac{b}{3}\mathrm{Tr}[\mathbf Q^3]+\frac{1}{4}(\mathrm{Tr}[\mathbf Q^2])^2
\end{equation}
We allow for normal fluctuations $\langle u_{\nu}^2\rangle$ in a nematic shell of constant thickness and add an anchoring free energy~\cite{Golovaty2015,Golovaty2017,Voigt2018} over a three-dimensional nematic tensor
\begin{align}\label{eq:f-anch}
f_{\rm anch}(\mathbf Q)&=\frac{w}{2}\left(\hat{\bm\nu}\cdot\mathbf Q\cdot\hat{\bm\nu}+\frac{1}{2}\right)^2 \\ \nonumber
&+\frac{w}{4}|\bm 1_{\mathcal S}\cdot(\hat{\bm\nu}\cdot\mathbf Q+\mathbf Q\cdot\hat{\bm\nu})|^2 \\ \nonumber
&=\frac{9w}{8}\left[\langle u_{\nu}^2\rangle+|\langle u_{\nu}\mathbf u_{\mathcal S}\rangle|^2\right] \\ \nonumber
&=\frac{w}{2}[(Q_{\nu\nu}+1/2)^2+Q_{\nu t_1}^2+Q_{\nu t_2}^2]
\end{align}
such that $f_{\rm tot}=f_{LdG}+f_{\rm anch}$ where $\bm 1_{\mathcal S}=\hat{\mathbf t}_u\hat{\mathbf t}_u+\hat{\mathbf t}_v\hat{\mathbf t}_v=\bm 1-\hat{\bm\nu}\hat{\bm\nu}$ is the tangential projector. Note that it differs from Ref.~\cite{Fournier2005} because it directly incorporates the knowledge in the statistical definition of $\mathbf Q$, instead of starting from its mean field expression and imposing a preferential boundary order $S_b$. Hence, the two terms in Eq.~\ref{eq:f-anch} ensure that out-of-plane orientations $u_{\nu}$ are penalized, such that $\mathbf Q$ prefers to be equal to $\mathbf Q_{\rm tangent}$ in the limit $w\rightarrow\infty$. For simplicity, we do not restrict the anchoring to a surface energy, and this bulk approximation should be valid for sufficiently thin shells with symmetric boundary conditions on 'top' and 'bottom'.

We can parametrize the $\mathbf Q$-tensor in the eigen-basis $(\hat{\mathbf n}_{\mathcal S},\hat{\mathbf n}_{\mathcal S}\times\hat{\bm\nu},\hat{\bm\nu})$ with two scalars $s=\langle(\hat{\mathbf u}\cdot\hat{\mathbf n}_{\mathcal S})^2\rangle=(1+S_{2D})/2$ and $\epsilon=\langle u_{\nu}^2\rangle$ such that $\mathbf Q=\frac{1}{2}\mathrm{Diag}[3s-1,2-3(s+\epsilon),3\epsilon-1]$. The anchoring free energy becomes $f_{\rm anch}=9w\epsilon^2/2$. Minimization of $f_{\rm tot}$ gives two coupled equations in $s$,$\epsilon$ which can be solved numerically as a function of three parameters $a$, $b$ and $w$.

\begin{figure}
\centering
\includegraphics[width=1.\linewidth]{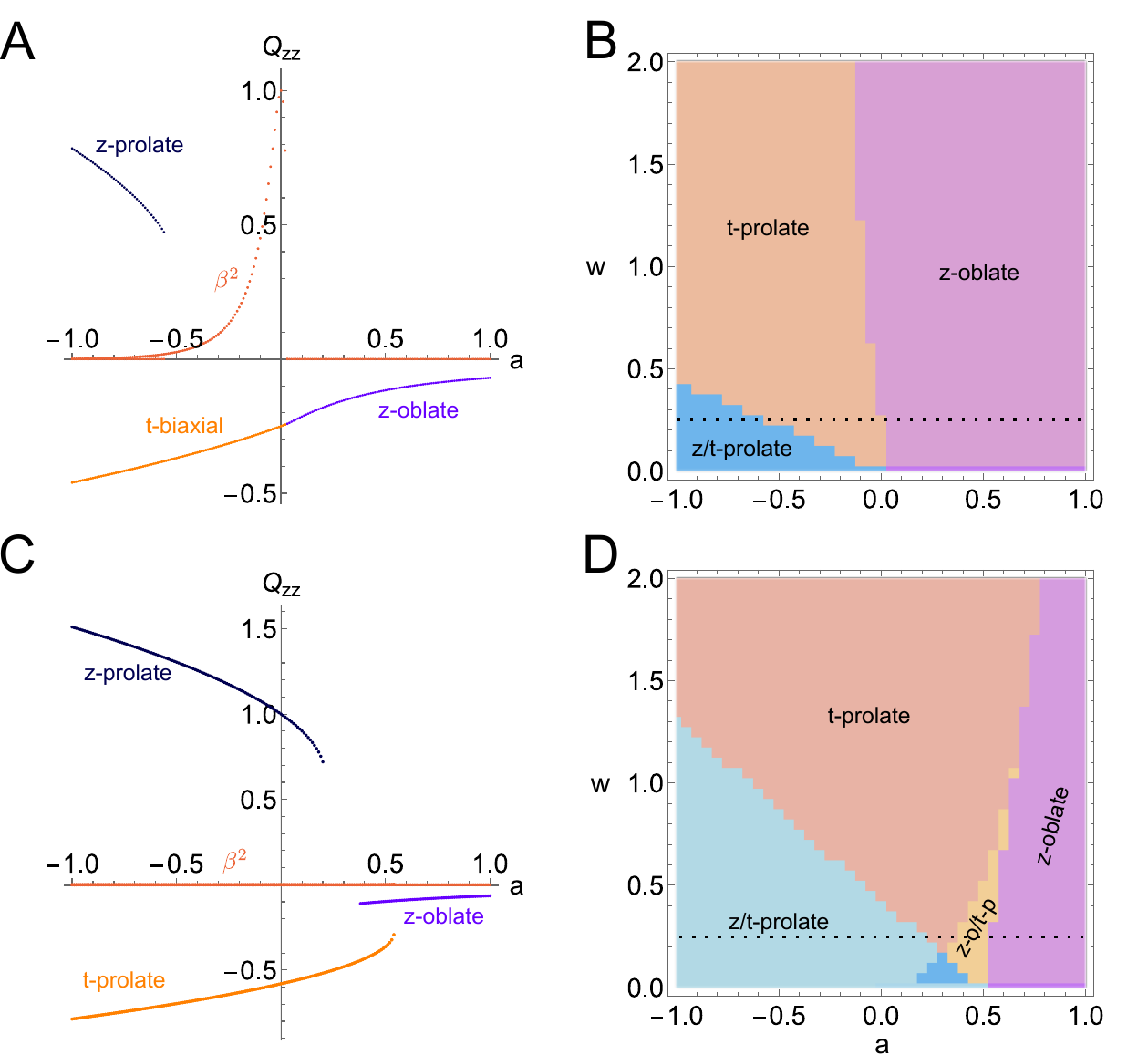}
\caption{Landau-de Gennes free energy with quadratic tangential anchoring.
(\textbf{A,C}) Normal nematic tensor component $Q_{\nu\nu}$ as a function of $a$ for $w=0.25$, with biaxial coefficient $\beta^2$ (red). $b=0.5$ (A), $b=3.5$ (C). (\textbf{B,D}) Phase diagram of the isotropic-nematic states in the parametric plane $(a,w)$. $b=0.5$ (B), $b=3.5$ (D).
}
\label{figS1}
\end{figure}

The results are shown on Fig.~\ref{figS1}B-D. When $a>0$, the tendency for an isotropic state $\mathbf Q=\bm 0$ at $w=0$ is continuously transformed into an oblate state (planar-isotropic) along $z$ as $w$ increases, because of the reduction of $u_{\nu}$ imposed by anchoring. When $a<0$, the tendency for a uniaxial nematic state ($w=0$) is balanced with the restriction of out-of-plane fluctuations as $w$ increases to form a biaxial state in the plane. Intriguingly, for sufficiently low values of $w$ and $a<0$, an additional prolate phase emerges either in the plane or along $z$. Here, the effect of anchoring is not strong enough and the material prefers to follow uniaxial configuration as in bulk nematics. For a small value of $b$ (Fig.~\ref{figS1}B), we get a second order phase transition as $a$ becomes negative.

For a large value of $b$ and sufficiently small $w$, the transition becomes first order, as shown on Fig.~\ref{figS1}C with $Q_z$ as a function of $a$ for fixed $w$. The z-oblate state at $a>0$ separates in two branches as $a$ decreases, the top one with $Q_z>0$ corresponding to z-prolate state, and the bottom one with $Q_z<0$ corresponding to t-prolate state (Fig.~\ref{figS1}C). This implies the possibility of phase coexistence between z-oblate (planar-isotropic) and t-prolate (planar-nematic) states. For larger values of $w$, the transition is second order again as expected from the limit $w\rightarrow\infty$ where two and three-dimensional descriptions become equivalent. Note also that for large values of $b$, biaxiality is essentially negligible and the t-biaxial state of Fig.~\ref{figS1}B becomes a t-prolate state in Fig.~\ref{figS1}D.

\section{Stiff anchoring energy and strong anchoring limit}\label{app:strong-anch}
\begin{figure}
\centering
\includegraphics[width=1.\linewidth]{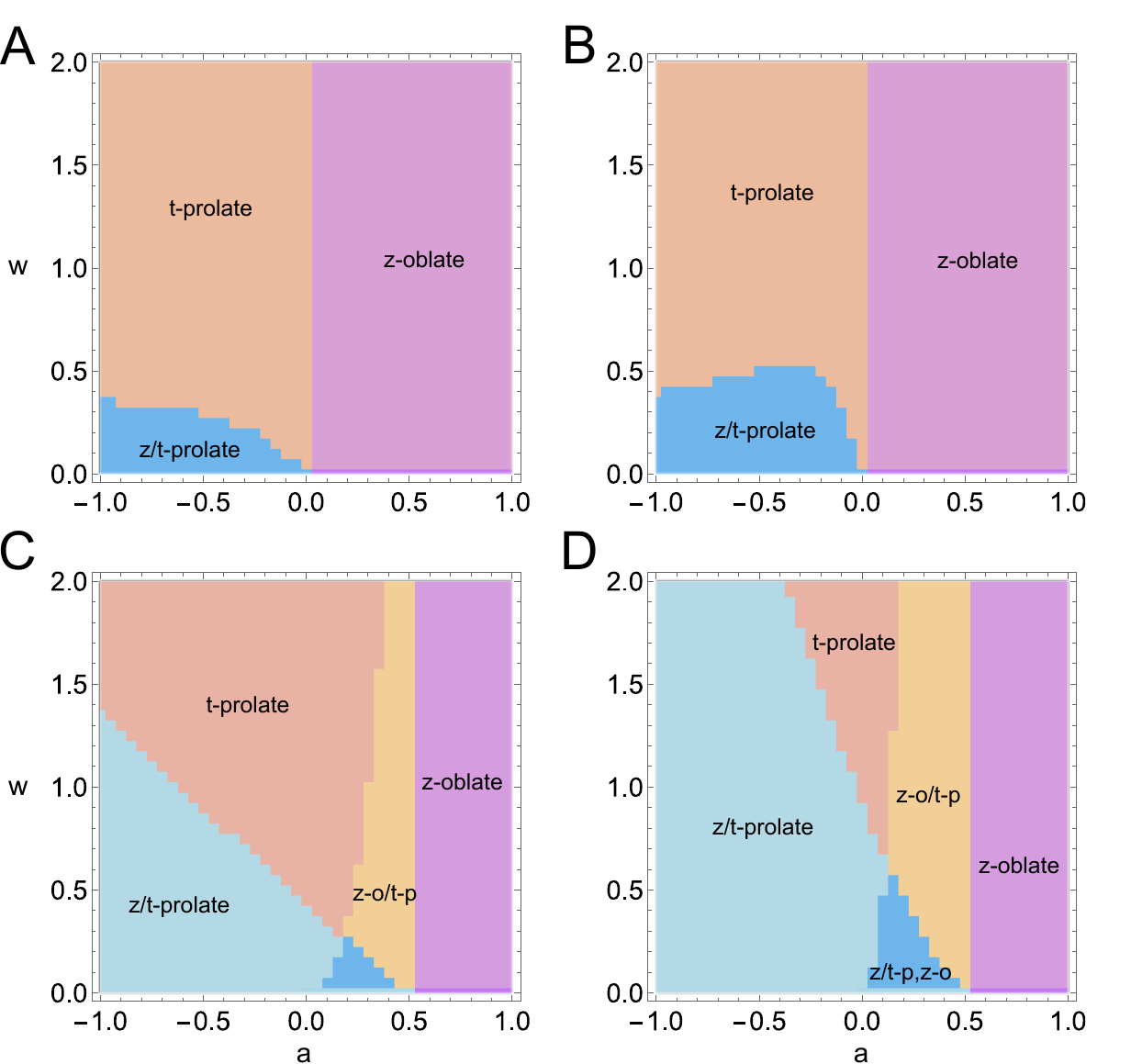}
\caption{Phase diagram of the Landau-de Gennes free energy with quartic (A,C) or hexatic (B,D) tangential anchoring, as a function of $a$ and $w$. The case $b=0.5$ is shown on (A,B), and the case $b=3.5$ on (C,D).}
\label{figS2}
\end{figure}

Here we replace the quadratic anchoring energy from Eq.~\ref{eq:f-anch} by a larger power $p>2$, such that
\begin{equation}\label{eq:f-anch-p}
f_{\rm anch}(\mathbf Q)=\frac{w}{2}[(Q_{\nu\nu}+1/2)^p+Q_{\nu t_1}^p+Q_{\nu t_2}^p]
\end{equation}
The anchoring free energy becomes $f_{\rm anch}=3^p\,w\epsilon^p/2$ for the previous parametrization of $\mathbf Q$.
For quartic ($p=4$) or hexatic ($p=6$) power, we compute the phase diagram from the minimization of $f_{\rm tot}=f_{LdG}+f_{\rm anch}$, represented on Fig.~\ref{figS2} for the values of $b$ used in Fig.~\ref{figS1}. One finds that a higher power extends the region where nematic and isotropic states coexist in the case $b=3.5$, a signature of a first order phase transition. For $b=0.5$, the coexistence region for the z-prolate decreases as the anchoring power increases.
\\\\
In the limit case of infinite anchoring strength $w\rightarrow\infty$ such that $Q_{\nu\nu}+1/2=Q_{\nu i}=0$, one writes a surface free energy to satisfy these constraints with $f_{\rm anch}=\Lambda_{\nu}(Q_{\nu\nu}+1/2)+\Lambda_t(Q_{uu}+Q_{vv}+Q_{\nu\nu})$. The variables $\Lambda_{\nu},\Lambda_t$ are Lagrange multipliers. The natural boundary condition at the substrate $n=0$ remains written as $K(\hat{\bm\nu}\cdot\bm\nabla)\mathbf Q=\partial_{\mathbf Q}f_{\rm anch}$, where now $\partial_{\mathbf Q}f_{\rm anch}=(\Lambda_{\nu}+\Lambda_t)\hat{\bm\nu}\hat{\bm\nu}+\Lambda_t(\hat{\mathbf e}_u\hat{\mathbf e}_u+\hat{\mathbf e}_v\hat{\mathbf e}_v)$. For the boundary condition to be well-defined and satisfy the traceless property of $\mathbf Q$, one requires $\Lambda_{\nu}+3\Lambda_t=0$. One identifies $-6\Lambda_t=w(1+2Q_{\nu\nu})=\Lambda$ as in the main text from Eq.~\ref{eq:qtangent}. Then, by injecting the constraints $Q_{\nu\nu}+1/2=Q_{\nu i}=0$ in the thin film dynamics of $Q_{\nu\nu}$, Eq.~\ref{eq:qtangent}, one obtains the condition
\begin{align}
\Lambda=&3KH[c_u^2(1+2Q_{uu})+c_v^2(1+2Q_{vv})] \\ \nonumber
&+\frac{3\chi}{2}Hc(\mathbf Q)-\frac{9H\Gamma\lambda}{2}(\bm\nabla_{\mathcal S}\cdot\mathbf v_{\mathcal S})
\end{align}
with $[\Delta\mathbf Q]_{\nu\nu}=c_u^2(1+2Q_{uu})+c_v^2(1+2Q_{vv})$. This is equivalent to Eq.~\ref{eq:Lambda} in the main text.

\begin{figure}
\centering
\includegraphics[width=1.\linewidth]{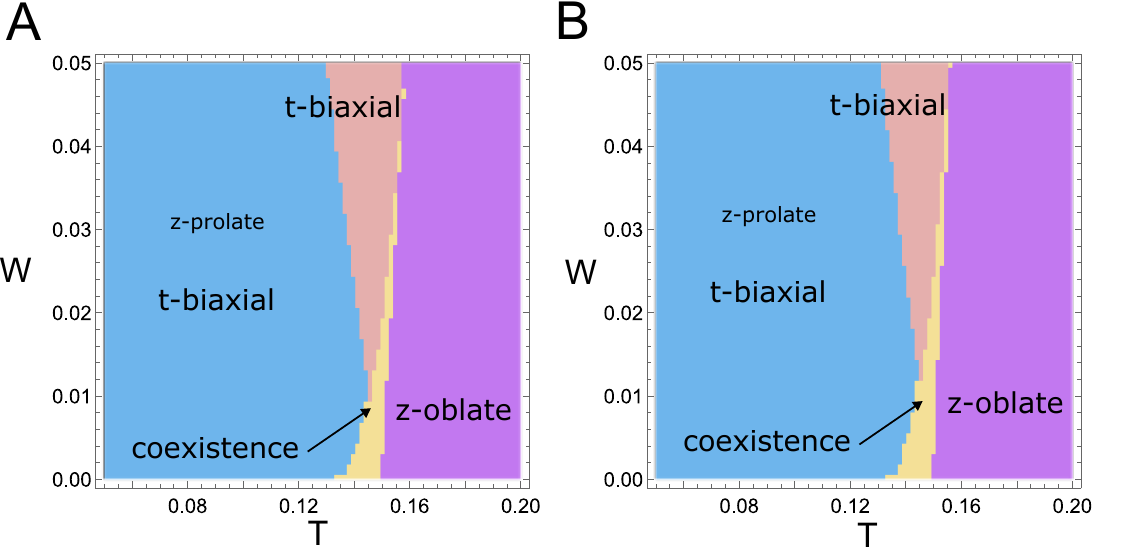}
\caption{Phase diagram of the Maier-Saupe free energy with quartic (A) or hexatic (B) tangential anchoring, as a function of $T$ and $W$.}
\label{figS3}
\end{figure}

\newpage
\section{Field gradients in curvilinear coordinates}\label{app:derivatives}
An orthonormal vector basis $\{\hat{\mathbf e}_u,\hat{\mathbf e}_v,\hat{\bm\nu}\}$ on the surfaces $\mathcal S_{n}$ is possible by aligning the tangent vectors with the principal curvatures. One obtains the metric tensor $\bm g_{n}=E_{n}\hat{\mathbf e}_u\hat{\mathbf e}_u+G_{n}\hat{\mathbf e}_v\hat{\mathbf e}_v$ with $E_{n}=E_0[1+n c_u]^2$, $G_{n}=G_0[1+n c_v]^2$. The curvature tensor is $\bm c_{n}=\bm\nabla\hat{\bm\nu}=\frac{c_u}{1+n c_u}\hat{\mathbf e}_u\hat{\mathbf e}_u+\frac{c_v}{1+n c_v}\hat{\mathbf e}_v\hat{\mathbf e}_v$. Then, one can show that the derivatives of basis vectors can be written as
\begin{align}
\partial_u\hat{\mathbf e}_u&=a_{uu}\hat{\mathbf e}_v-c_u\sqrt{E_0}\hat{\bm\nu},\quad
\partial_v\hat{\mathbf e}_u=-a_{vv}\hat{\mathbf e}_v \\ \nonumber
\partial_v\hat{\mathbf e}_v&=a_{vv}\hat{\mathbf e}_u-c_v\sqrt{G_0}\hat{\bm\nu},\quad
\partial_u\hat{\mathbf e}_v=-a_{uu}\hat{\mathbf e}_u \\ \nonumber
\partial_u\hat{\bm\nu}&=c_u\sqrt{E_0}\hat{\mathbf e}_u,\quad
\partial_v\hat{\bm\nu}=c_v\sqrt{G_0}\hat{\mathbf e}_v
\end{align}
where
\begin{align}\label{eq:aii}
a_{uu}&=-\frac{\partial_vE_0[1+n c_u]+2n E_0\partial_vc_u}{2\sqrt{E_0G_0}[1+n c_v]} \\ \nonumber
a_{vv}&=-\frac{\partial_uG_0[1+n c_v]+2G_0n\partial_uc_v}{2\sqrt{E_0G_0}[1+n c_u]}
\end{align}
With those transformation rules, one can compute the pressure gradient
\begin{align}
\bm\nabla P=\frac{\hat{\mathbf e}_u\partial_uP}{\sqrt{E_0}[1+n c_u]}+\frac{\hat{\mathbf e}_v\partial_vP}{\sqrt{G_0}[1+n c_v]}+\hat{\bm\nu}\partial_{n}P
\end{align}
the flow divergence
\begin{align}
\bm\nabla\cdot\mathbf v&=\frac{\partial_uv_u-a_{uu}v_v}{\sqrt{E_0}[1+n c_u]}+\frac{\partial_vv_v-a_{vv}v_u}{\sqrt{G_0}[1+n c_v]}+\partial_{n}v_{\nu} \\ \nonumber
&+\left(\frac{c_u}{1+n c_u}+\frac{c_v}{1+n c_v}\right)v_{\nu}
\end{align}
the vorticity
\begin{align}
\bm\omega&= \\ \nonumber
&\left[\frac{\partial_uv_v+a_{uu}v_u}{2\sqrt{E_{n}}}-\frac{\partial_vv_u+a_{vv}v_v}{2\sqrt{G_{n}}}\right](\hat{\mathbf e}_u\hat{\mathbf e}_v-\hat{\mathbf e}_v\hat{\mathbf e}_u) \\ \nonumber
&+\frac{1}{2}\left[\frac{c_u v_u}{1+n c_u}-\frac{\partial_uv_{\nu}}{\sqrt{E_{n}}}+\partial_{n}v_u\right](\hat{\bm\nu}\hat{\mathbf e}_u-\hat{\mathbf e}_u\hat{\bm\nu}) \\ \nonumber
&+\frac{1}{2}\left[\frac{c_vv_v}{1+n c_v}-\frac{\partial_vv_{\nu}}{\sqrt{G_{n}}}+\partial_{n}v_v\right](\hat{\bm\nu}\hat{\mathbf e}_v-\hat{\mathbf e}_v\hat{\bm\nu}) \\ \nonumber
&=\omega_{uv}(\hat{\mathbf e}_u\hat{\mathbf e}_v-\hat{\mathbf e}_v\hat{\mathbf e}_u)+\omega_{\nu i}(\hat{\bm\nu}\hat{\mathbf e}_i-\hat{\mathbf e}_i\hat{\bm\nu})
\end{align}
and the symmetric strain rate
\begin{align}
\tilde{\mathbf u}&= \\ \nonumber
&+\left[\frac{\partial_uv_u-a_{uu}v_v}{\sqrt{E_{n}}}+\frac{c_uv_{\nu}}{1+n c_u}\right]\hat{\mathbf e}_u\hat{\mathbf e}_u \\ \nonumber
&+\left[\frac{\partial_vv_v-a_{vv}v_u}{\sqrt{G_{n}}}+\frac{c_vv_{\nu}}{1+n c_v}\right]\hat{\mathbf e}_v\hat{\mathbf e}_v \\ \nonumber
&+\partial_{n}v_{\nu}\hat{\bm\nu}\hat{\bm\nu} \\ \nonumber
&+\left[\frac{\partial_uv_v+a_{uu}v_u}{2\sqrt{E_{n}}}+\frac{\partial_vv_u+a_{vv}v_v}{2\sqrt{G_{n}}}\right](\hat{\mathbf e}_u\hat{\mathbf e}_v+\hat{\mathbf e}_v\hat{\mathbf e}_u) \\ \nonumber
&+\frac{1}{2}\left[\partial_{n}v_u+\frac{\partial_uv_{\nu}}{\sqrt{E_{n}}}-\frac{c_uv_u}{1+n c_u}\right](\hat{\mathbf e}_u\hat{\bm\nu}+\hat{\bm\nu}\hat{\mathbf e}_u) \\ \nonumber
&+\frac{1}{2}\left[\partial_{n}v_v+\frac{\partial_vv_{\nu}}{\sqrt{G_{n}}}-\frac{c_vv_v}{1+n c_v}\right](\hat{\mathbf e}_v\hat{\bm\nu}+\hat{\bm\nu}\hat{\mathbf e}_v)
\end{align}
In addition, one needs to compute the nematic gradient tensor $\bm\nabla\mathbf Q$. We write the nematic tensor as
\begin{align}
\mathbf Q&=Q_{uu}\hat{\mathbf e}_u\hat{\mathbf e}_u-(Q_{uu}+Q_{\nu\nu})\hat{\mathbf e}_v\hat{\mathbf e}_v+Q_{\nu\nu}\hat{\bm\nu}\hat{\bm\nu} \\ \nonumber
&+Q_{uv}(\hat{\mathbf e}_u\hat{\mathbf e}_v+\hat{\mathbf e}_v\hat{\mathbf e}_u) \\ \nonumber
&+Q_{\nu u}(\hat{\bm\nu}\hat{\mathbf e}_u+\hat{\mathbf e}_u\hat{\bm\nu})+Q_{\nu v}(\hat{\bm\nu}\hat{\mathbf e}_v+\hat{\mathbf e}_v\hat{\bm\nu})
\end{align}
with $Q_{vv}=-Q_{uu}-Q_{\nu\nu}$ and get
\begin{align}
&\bm\nabla\mathbf Q= \\ \nonumber
&+\hat{\mathbf e}_u\left[\frac{\partial_uQ_{uu}-2a_{uu}Q_{uv}}{\sqrt{E_n}}+\frac{2c_uQ_{\nu u}}{1+nc_u}\right]\hat{\mathbf e}_u\hat{\mathbf e}_u \\ \nonumber
&+\hat{\mathbf e}_u\left[\frac{\partial_uQ_{uv}+2a_{uu}(2Q_{uu}+Q_{\nu\nu})}{\sqrt{E_n}}+\frac{c_uQ_{\nu v}}{1+nc_u}\right](\hat{\mathbf e}_u\hat{\mathbf e}_v+\hat{\mathbf e}_v\hat{\mathbf e}_u) \\ \nonumber
&+\hat{\mathbf e}_u\left[\frac{-\partial_uQ_{uu}-\partial_uQ_{\nu\nu}+2a_{uu}Q_{uv}}{\sqrt{E_n}}\right]\hat{\mathbf e}_v\hat{\mathbf e}_v \\ \nonumber
&+\hat{\mathbf e}_u\left[\frac{\partial_uQ_{\nu u}-2a_{uu}Q_{\nu v}}{\sqrt{E_n}}+\frac{c_u(Q_{\nu\nu}-Q_{uu})}{1+nc_u}\right](\hat{\mathbf e}_u\hat{\bm\nu}+\hat{\bm\nu}\hat{\mathbf e}_u) \\ \nonumber
&+\hat{\mathbf e}_u\left[\frac{\partial_uQ_{\nu v}+2a_{uu}Q_{\nu u}}{\sqrt{E_n}}-\frac{c_uQ_{uv}}{1+nc_u}\right](\hat{\mathbf e}_v\hat{\bm\nu}+\hat{\bm\nu}\hat{\mathbf e}_v) \\ \nonumber
&+\hat{\mathbf e}_u\left[\frac{\partial_uQ_{\nu\nu}}{\sqrt{E_n}}-\frac{2c_uQ_{\nu u}}{1+nc_u}\right]\hat{\bm\nu}\hat{\bm\nu} \\ \nonumber
&+\hat{\mathbf e}_v\left[\frac{\partial_vQ_{uu}+2a_{vv}Q_{uv}}{\sqrt{G_n}}\right]\hat{\mathbf e}_u\hat{\mathbf e}_u \\ \nonumber
&+\hat{\mathbf e}_v\left[\frac{\partial_vQ_{uv}-2a_{vv}(2Q_{uu}+Q_{\nu\nu})}{\sqrt{G_n}}+\frac{c_vQ_{\nu u}}{1+nc_v}\right](\hat{\mathbf e}_u\hat{\mathbf e}_v+\hat{\mathbf e}_v\hat{\mathbf e}_u) \\ \nonumber
&+\hat{\mathbf e}_v\left[\frac{-\partial_vQ_{uu}-\partial_vQ_{\nu\nu}-2a_{vv}Q_{uv}}{\sqrt{G_n}}+\frac{2c_vQ_{\nu v}}{1+nc_v}\right]\hat{\mathbf e}_v\hat{\mathbf e}_v \\ \nonumber
&+\hat{\mathbf e}_v\left[\frac{\partial_vQ_{\nu u}+2a_{vv}Q_{\nu v}}{\sqrt{G_n}}-\frac{c_vQ_{uv}}{1+nc_v}\right](\hat{\mathbf e}_u\hat{\bm\nu}+\hat{\bm\nu}\hat{\mathbf e}_u) \\ \nonumber
&+\hat{\mathbf e}_v\left[\frac{\partial_vQ_{\nu v}-2a_{vv}Q_{\nu u}}{\sqrt{G_n}}+\frac{c_v(Q_{uu}+2Q_{\nu\nu})}{1+nc_v}\right](\hat{\mathbf e}_v\hat{\bm\nu}+\hat{\bm\nu}\hat{\mathbf e}_v) \\ \nonumber
&+\hat{\mathbf e}_v\left[\frac{\partial_vQ_{\nu\nu}}{\sqrt{G_n}}-\frac{2c_vQ_{\nu v}}{1+nc_v}\right]\hat{\bm\nu}\hat{\bm\nu} \\ \nonumber
&+\hat{\bm\nu}\,\partial_{n}\mathbf Q
\end{align}

\section{Free surface details}\label{app:free-surface}
From the surface function $F(u,v,n,t)=H(u,v,t)-n$, one obtains the following kinematic condition
\begin{align}
\partial_tH+\frac{v_u}{\sqrt{E_{n}}}\partial_uH+\frac{v_v}{\sqrt{G_{n}}}\partial_vH-v_{\nu}=0
\end{align}
at $n=H$. The normal to the interface can be computed explicitly from $\bm N=(\bm B_u\times\bm B_v)/|\bm B_u\times\bm B_v|$ with
\begin{align}
\bm N=\frac{-\sqrt{G_{n}}\partial_uH\hat{\mathbf e}_u-\sqrt{E_{n}}\partial_vH\hat{\mathbf e}_v+\sqrt{E_{n}G_{n}}\hat{\bm\nu}}{\sqrt{G_{n}(\partial_uH)^2+E_{n}(\partial_vH)^2+E_{n}G_{n}}}
\end{align}
leading to $\bm N^{(0)}=\hat{\bm\nu}$ and $\bm N^{(1)}=-\bm\nabla_{\mathcal S}H=-(\partial_uH/\sqrt{E_0})\hat{\mathbf e}_u-(\partial_vH/\sqrt{G_0})\hat{\mathbf e}_v$. Similarly, $\bm B_i^{(0)}=\mathbf b_i(0)$ and $\bm B_i^{(1)}=H c_i\mathbf b_i(0)+\partial_iH\,\hat{\bm\nu}$.

At zero-th order, the interface curvature components are $C_{ij}^{(0)}=\mathbf b_i(0)\cdot\partial_j\hat{\bm\nu}=\mathrm{Diag}[c_u E_0,c_v G_0]$ and the principal curvatures follow from $\mathbf C^{(0)}=C_{ij}^{(0)}\mathbf b^i(0)\mathbf b^j(0)=c_u\hat{\mathbf e}_u\hat{\mathbf e}_u+c_v\hat{\mathbf e}_v\hat{\mathbf e}_v$. Hence, one finds $C_1^{(0)}=c_u$ and $C_2^{(0)}=c_v$.

In absence of substrate curvature, $c_u=c_v=0$, one has $E_n=G_n=1$ in Cartesian coordinates and the curvature tensor becomes at least order $\mathbf C^{(1)}=-\partial_{ij}H\hat{\mathbf e}_i\hat{\mathbf e}_j$. One gets $C_1^{(1)}+C_2^{(1)}=-\Delta H$.

\section{Expansion of boundary conditions}\label{app:bcs}
The boundary conditions at the substrate $n=0$ can be expressed as the following hierarchy up to $\mathcal{O}(\epsilon)$
\begin{align}
\hat{\bm\nu}\cdot\bm\sigma\cdot\hat{\mathbf e}_i&=\xi_s^{(1)}v_i^{(0)} \\ \nonumber
&=\eta\partial_{n}v_i^{(0)} \\ \nonumber
&+\eta[\partial_{n}v_i^{(1)}-c_iv_i^{(0)}]+\alpha Q_{\nu i}^{(0)} \\ \nonumber
&+\eta[\partial_{n}v_i^{(2)}-c_iv_i^{(1)}]+\alpha Q_{\nu i}^{(1)} \\
K(\hat{\bm\nu}\cdot\bm\nabla)\mathbf Q&=\frac{1}{2}w^{(1)}[(1+2Q_{\nu\nu}^{(0)})\hat{\bm\nu}\hat{\bm\nu}+Q_{i\nu}^{(0)}(\hat{\mathbf e}_i\hat{\bm\nu}+\hat{\mathbf e}_i\hat{\bm\nu})] \\ \nonumber
&=K\partial_{n}\mathbf Q^{(0)} \\ \nonumber
&+K\partial_{n}\mathbf Q^{(1)} \\ \nonumber
&+K\partial_{n}\mathbf Q^{(2)}
\end{align}
where $\xi_s^{(1)}$ and $w^{(1)}$ indicate that those parameters scale as $\mathcal{O}(\epsilon)$.

At the free interface, one obtains the following hierarchy of stress boundary conditions, $\bm N\cdot\bm\sigma\cdot\bm B_i=0$ up to order $\mathcal{O}(\epsilon)$ and $\bm N\cdot\bm\sigma\cdot\bm N=-\gamma(C_1+C_2)$ up to order $\mathcal{O}(1)$, respectively
\begin{align}
\bm N\cdot\bm\sigma\cdot\frac{\bm B_i}{|\mathbf b_i(0)|}&=0 \\ \nonumber
&=\eta\partial_{n}v_i^{(0)} \\ \nonumber
&+\eta[\partial_{n}v_i^{(1)}-c_iv_i^{(0)}]+\alpha Q_{\nu i}^{(0)} \\ \nonumber
&+c_i H[\eta\partial_{n}v_i^{(1)}+\alpha Q_{\nu i}^{(0)}] \\ \nonumber
&+\eta[\partial_{n}v_i^{(2)}+\bm\nabla_{\mathcal S,i}v_{\nu}^{(1)}-c_iv_i^{(1)}]+\alpha Q_{\nu i}^{(1)} \\ \nonumber
&-\nabla_{\mathcal S,j}H[-P^{(0)}\delta_{ij}+2\eta\tilde{u}_{ij}^{(0)}+\alpha Q_{ij}^{(0)}] \\ \nonumber
&+\nabla_{\mathcal S,i}H\,[-P^{(0)}+2\eta\partial_{n}v_{\nu}^{(1)}+\alpha Q_{\nu\nu}^{(0)}] \\
\bm N\cdot\bm\sigma\cdot\bm N&=-\gamma(C_1^{(0)}+C_2^{(0)}) \\ \nonumber
&=-P^{(0)}+2\eta\partial_{n}v_{\nu}^{(1)}+\alpha Q_{\nu\nu}^{(0)}
\end{align}
In addition, nematic free anchoring gives
\begin{align}
(\bm N\cdot\bm\nabla)\mathbf Q&=0 \\ \nonumber
&=\partial_{n}\mathbf Q^{(0)} \\ \nonumber
&+\partial_{n}\mathbf Q^{(1)} \\ \nonumber
&+\partial_{n}\mathbf Q^{(2)}-\nabla_{\mathcal S,j}H\,\hat{\mathbf e}_j\cdot[\bm\nabla\mathbf Q]^{(0)}
\end{align}
at order $\mathcal{O}(\epsilon)$.

\begin{widetext}
\section{Gradient formula at zero-th order}\label{app:derivatives0}
In this section, we present formulas used in final equations, Eq.~\ref{eq:qtangent} and Eq.~\ref{eq:ftangent}. Quantities of order $\mathcal{O}(1)$ are written without explicit superscript for clarity.

The tangential gradient of pressure and height are
\begin{align}
\bm\nabla_{\mathcal S}P^{(0)}=\frac{\hat{\mathbf e}_u\partial_uP}{\sqrt{E_0}}+\frac{\hat{\mathbf e}_v\partial_vP}{\sqrt{G_0}},\qquad
\bm\nabla_{\mathcal S}H=\frac{\hat{\mathbf e}_u\partial_uH}{\sqrt{E_0}}+\frac{\hat{\mathbf e}_v\partial_vH}{\sqrt{G_0}}
\end{align}
The vorticity is
\begin{align}
\bm\omega^{(0)}=
\left[\frac{\partial_uv_v+a_{uu}v_u}{2\sqrt{E_0}}-\frac{\partial_vv_u+a_{vv}v_v}{2\sqrt{G_0}}\right](\hat{\mathbf e}_u\hat{\mathbf e}_v-\hat{\mathbf e}_v\hat{\mathbf e}_u)
+\frac{1}{2}\left[c_iv_i+\partial_{n}v_i^{(1)}\right](\hat{\bm\nu}\hat{\mathbf e}_i-\hat{\mathbf e}_i\hat{\bm\nu})
\end{align}
The symmetric strain rate is
\begin{align}
\tilde{\mathbf u}^{(0)}&=\frac{\partial_uv_u-a_{uu}v_v}{\sqrt{E_0}}\hat{\mathbf e}_u\hat{\mathbf e}_u+\frac{\partial_vv_v-a_{vv}v_u}{\sqrt{G_0}}\hat{\mathbf e}_v\hat{\mathbf e}_v
+\partial_{n}v_{\nu}^{(1)}\hat{\bm\nu}\hat{\bm\nu}
\\ \nonumber
&+\left[\frac{\partial_uv_v+a_{uu}v_u}{2\sqrt{E_0}}+\frac{\partial_vv_u+a_{vv}v_v}{2\sqrt{G_0}}\right](\hat{\mathbf e}_u\hat{\mathbf e}_v+\hat{\mathbf e}_v\hat{\mathbf e}_u)
+\frac{1}{2}[\partial_{n}v_i^{(1)}-c_iv_i](\hat{\mathbf e}_i\hat{\bm\nu}+\hat{\bm\nu}\hat{\mathbf e}_i)
\end{align}
The vectorial Laplacian projected on the tangent plane is
\begin{align}\label{eq:laplacian-v}
\bm 1_{\mathcal S}\cdot[\Delta\mathbf v]^{(0)}&=
\\ \nonumber
&\frac{1}{E_0}([\partial_u^2v_u-2a_{uu}\partial_uv_v-v_v\partial_ua_{uu}]\hat{\mathbf e}_u
+[\partial_u^2v_v+2a_{uu}\partial_uv_u+v_u\partial_ua_{uu}]\hat{\mathbf e}_v) \\ \nonumber
&+\frac{1}{G_0}(
[\partial_v^2v_u+2a_{vv}\partial_vv_v+v_v\partial_va_{vv}]\hat{\mathbf e}_u
+[\partial_v^2v_v-2a_{vv}\partial_vv_u-v_u\partial_va_{vv}]\hat{\mathbf e}_v)
\\ \nonumber
&-\frac{a_{uu}^2}{E_0}(v_u\hat{\mathbf e}_u+v_v\hat{\mathbf e}_v)
-\frac{a_{vv}^2}{G_0}(v_u\hat{\mathbf e}_u+v_v\hat{\mathbf e}_v)
\\ \nonumber
&-\frac{1}{\sqrt{E_0G_0}}[a_{uu}\partial_vv_u+a_{vv}\partial_uv_u]\hat{\mathbf e}_u
-\frac{1}{\sqrt{E_0G_0}}[a_{uu}\partial_vv_v+a_{vv}\partial_uv_v]\hat{\mathbf e}_v
\\ \nonumber
&+[\partial_{n}^2v_i^{(2)}+(c_u+c_v)\partial_nv_i^{(1)}-c_i^2v_i]\hat{\mathbf e}_i
\end{align}
The nematic tensor gradient is
\begin{align}
&[\bm\nabla\mathbf Q]^{(0)}= \\ \nonumber
&+\hat{\mathbf e}_u\left[\frac{\partial_uQ_{uu}-2a_{uu}Q_{uv}}{\sqrt{E_0}}+2c_uQ_{\nu u}\right]\hat{\mathbf e}_u\hat{\mathbf e}_u
+\hat{\mathbf e}_v\left[\frac{\partial_vQ_{uu}+2a_{vv}Q_{uv}}{\sqrt{G_0}}\right]\hat{\mathbf e}_u\hat{\mathbf e}_u
\\ \nonumber
&+\hat{\mathbf e}_u\left[\frac{\partial_uQ_{uv}+2a_{uu}(2Q_{uu}+Q_{\nu\nu})}{\sqrt{E_0}}+c_uQ_{\nu v}\right](\hat{\mathbf e}_u\hat{\mathbf e}_v+\hat{\mathbf e}_v\hat{\mathbf e}_u)
+\hat{\mathbf e}_v\left[\frac{\partial_vQ_{uv}-2a_{vv}(2Q_{uu}+Q_{\nu\nu})}{\sqrt{G_0}}+c_vQ_{\nu u}\right](\hat{\mathbf e}_u\hat{\mathbf e}_v+\hat{\mathbf e}_v\hat{\mathbf e}_u)
\\ \nonumber
&+\hat{\mathbf e}_u\left[\frac{-\partial_uQ_{uu}-\partial_uQ_{\nu\nu}+2a_{uu}Q_{uv}}{\sqrt{E_0}}\right]\hat{\mathbf e}_v\hat{\mathbf e}_v
+\hat{\mathbf e}_v\left[\frac{-\partial_vQ_{uu}-\partial_vQ_{\nu\nu}-2a_{vv}Q_{uv}}{\sqrt{G_0}}+2c_vQ_{\nu v}\right]\hat{\mathbf e}_v\hat{\mathbf e}_v
\\ \nonumber
&+\hat{\mathbf e}_u\left[\frac{\partial_uQ_{\nu u}-2a_{uu}Q_{\nu v}}{\sqrt{E_0}}+c_u(Q_{\nu\nu}-Q_{uu})\right](\hat{\mathbf e}_u\hat{\bm\nu}+\hat{\bm\nu}\hat{\mathbf e}_u)
+\hat{\mathbf e}_v\left[\frac{\partial_vQ_{\nu u}+2a_{vv}Q_{\nu v}}{\sqrt{G_0}}-c_vQ_{uv}\right](\hat{\mathbf e}_u\hat{\bm\nu}+\hat{\bm\nu}\hat{\mathbf e}_u)
\\ \nonumber
&+\hat{\mathbf e}_u\left[\frac{\partial_uQ_{\nu v}+2a_{uu}Q_{\nu u}}{\sqrt{E_0}}-c_uQ_{uv}\right](\hat{\mathbf e}_v\hat{\bm\nu}+\hat{\bm\nu}\hat{\mathbf e}_v)
+\hat{\mathbf e}_v\left[\frac{\partial_vQ_{\nu v}-2a_{vv}Q_{\nu u}}{\sqrt{G_0}}+c_v(Q_{uu}+2Q_{\nu\nu})\right](\hat{\mathbf e}_v\hat{\bm\nu}+\hat{\bm\nu}\hat{\mathbf e}_v)
\\ \nonumber
&+\hat{\mathbf e}_u\left[\frac{\partial_uQ_{\nu\nu}}{\sqrt{E_0}}-2c_uQ_{\nu u}\right]\hat{\bm\nu}\hat{\bm\nu}
+\hat{\mathbf e}_v\left[\frac{\partial_vQ_{\nu\nu}}{\sqrt{G_0}}-2c_vQ_{\nu v}\right]\hat{\bm\nu}\hat{\bm\nu}
+\hat{\bm\nu}\partial_n\mathbf Q^{(1)}
\end{align}
The nematic divergence is
\begin{align}
[\bm\nabla\cdot\mathbf Q]^{(0)}&=
\left[\frac{\partial_uQ_{uu}-2a_{uu}Q_{uv}}{\sqrt{E_0}}+\frac{\partial_vQ_{uv}-2a_{vv}(2Q_{uu}+Q_{\nu\nu})}{\sqrt{G_0}}+(2c_u+c_v)Q_{\nu u}+\partial_nQ_{\nu u}^{(1)}\right]\hat{\mathbf e}_u
\\ \nonumber
&+\left[\frac{\partial_uQ_{uv}+2a_{uu}(2Q_{uu}+Q_{\nu\nu})}{\sqrt{E_0}}-\frac{\partial_vQ_{uu}+\partial_vQ_{\nu\nu}+2a_{vv}Q_{uv}}{\sqrt{G_0}}+(c_u+2c_v)Q_{\nu v}+\partial_nQ_{\nu v}^{(1)}\right]\hat{\mathbf e}_v
\\ \nonumber
&+\left[\frac{\partial_uQ_{\nu u}-2a_{uu}Q_{\nu v}}{\sqrt{E_0}}+\frac{\partial_vQ_{\nu v}-2a_{vv}Q_{\nu u}}{\sqrt{G_0}}+c_u(Q_{\nu\nu}-Q_{uu})+c_v(Q_{uu}+2Q_{\nu\nu})+\partial_nQ_{\nu\nu}^{(1)}\right]\hat{\bm\nu}
\end{align}
The nematic Laplacian is
\begin{align}
&[\Delta\mathbf Q]^{(0)}= \\ \nonumber
&+\partial_u\left[\frac{\partial_uQ_{uu}-2a_{uu}Q_{uv}}{\sqrt{E_0}}+2c_uQ_{\nu u}\right]\frac{\hat{\mathbf e}_u\hat{\mathbf e}_u}{\sqrt{E_0}}+\partial_v\left[\frac{\partial_vQ_{uu}+2a_{vv}Q_{uv}}{\sqrt{G_0}}\right]\frac{\hat{\mathbf e}_u\hat{\mathbf e}_u}{\sqrt{G_0}}
\\ \nonumber
&+\left[\frac{\partial_uQ_{uu}-2a_{uu}Q_{uv}}{\sqrt{E_0}}+2c_uQ_{\nu u}\right]\left[\frac{\partial_u(\hat{\mathbf e}_u\hat{\mathbf e}_u)}{\sqrt{E_0}}-\frac{a_{vv}\hat{\mathbf e}_u\hat{\mathbf e}_u}{\sqrt{G_0}}\right]+\left[\frac{\partial_vQ_{uu}+2a_{vv}Q_{uv}}{\sqrt{G_0}}\right]\left[\frac{\partial_v(\hat{\mathbf e}_u\hat{\mathbf e}_u)}{\sqrt{G_0}}-\frac{a_{uu}\hat{\mathbf e}_u\hat{\mathbf e}_u}{\sqrt{E_0}}\right]
\\ \nonumber
&+\partial_u\left[\frac{\partial_uQ_{uv}+2a_{uu}(2Q_{uu}+Q_{\nu\nu})}{\sqrt{E_0}}+c_uQ_{\nu v}\right]\frac{\hat{\mathbf e}_u\hat{\mathbf e}_v+\hat{\mathbf e}_v\hat{\mathbf e}_u}{\sqrt{E_0}}
+\partial_v\left[\frac{\partial_vQ_{uv}-2a_{vv}(2Q_{uu}+Q_{\nu\nu})}{\sqrt{G_0}}+c_vQ_{\nu u}\right]\frac{\hat{\mathbf e}_u\hat{\mathbf e}_v+\hat{\mathbf e}_v\hat{\mathbf e}_u}{\sqrt{G_0}}
\\ \nonumber
&+\left[\frac{\partial_uQ_{uv}+2a_{uu}(2Q_{uu}+Q_{\nu\nu})}{\sqrt{E_0}}+c_uQ_{\nu v}\right]
\left[\frac{\partial_u(\hat{\mathbf e}_u\hat{\mathbf e}_v+\hat{\mathbf e}_v\hat{\mathbf e}_u)}{\sqrt{E_0}}-\frac{a_{vv}(\hat{\mathbf e}_u\hat{\mathbf e}_v+\hat{\mathbf e}_v\hat{\mathbf e}_u)}{\sqrt{G_0}}\right]
\\ \nonumber
&+\left[\frac{\partial_vQ_{uv}-2a_{vv}(2Q_{uu}+Q_{\nu\nu})}{\sqrt{G_0}}+c_vQ_{\nu u}\right]\left[\frac{\partial_v(\hat{\mathbf e}_u\hat{\mathbf e}_v+\hat{\mathbf e}_v\hat{\mathbf e}_u)}{\sqrt{G_0}}-\frac{a_{uu}(\hat{\mathbf e}_u\hat{\mathbf e}_v+\hat{\mathbf e}_v\hat{\mathbf e}_u)}{\sqrt{E_0}}\right]
\\ \nonumber
&+\partial_u\left[\frac{-\partial_uQ_{uu}-\partial_uQ_{\nu\nu}+2a_{uu}Q_{uv}}{\sqrt{E_0}}\right]\frac{\hat{\mathbf e}_v\hat{\mathbf e}_v}{\sqrt{E_0}}
+\partial_v\left[\frac{-\partial_vQ_{uu}-\partial_vQ_{\nu\nu}-2a_{vv}Q_{uv}}{\sqrt{G_0}}+2c_vQ_{\nu v}\right]\frac{\hat{\mathbf e}_v\hat{\mathbf e}_v}{\sqrt{G_0}}
\\ \nonumber
&+\left[\frac{-\partial_uQ_{uu}-\partial_uQ_{\nu\nu}+2a_{uu}Q_{uv}}{\sqrt{E_0}}\right]\left[\frac{\partial_u(\hat{\mathbf e}_v\hat{\mathbf e}_v)}{\sqrt{E_0}}-\frac{a_{vv}\hat{\mathbf e}_v\hat{\mathbf e}_v}{\sqrt{G_0}}\right]
\\ \nonumber
&+\left[\frac{-\partial_vQ_{uu}-\partial_vQ_{\nu\nu}-2a_{vv}Q_{uv}}{\sqrt{G_0}}+2c_vQ_{\nu v}\right]\left[\frac{\partial_v(\hat{\mathbf e}_v\hat{\mathbf e}_v)}{\sqrt{G_0}}-\frac{a_{uu}\hat{\mathbf e}_v\hat{\mathbf e}_v}{\sqrt{E_0}}\right]
\\ \nonumber
&+\partial_u\left[\frac{\partial_uQ_{\nu u}-2a_{uu}Q_{\nu v}}{\sqrt{E_0}}+c_u(Q_{\nu\nu}-Q_{uu})\right]\frac{\hat{\mathbf e}_u\hat{\bm\nu}+\hat{\bm\nu}\hat{\mathbf e}_u}{\sqrt{E_0}}
+\partial_v\left[\frac{\partial_vQ_{\nu u}+2a_{vv}Q_{\nu v}}{\sqrt{G_0}}-c_vQ_{uv}\right]\frac{\hat{\mathbf e}_u\hat{\bm\nu}+\hat{\bm\nu}\hat{\mathbf e}_u}{\sqrt{G_0}}
\\ \nonumber
&+\left[\frac{\partial_uQ_{\nu u}-2a_{uu}Q_{\nu v}}{\sqrt{E_0}}+c_u(Q_{\nu\nu}-Q_{uu})\right]
\left[\frac{\partial_u(\hat{\mathbf e}_u\hat{\bm\nu}+\hat{\bm\nu}\hat{\mathbf e}_u)}{\sqrt{E_0}}-\frac{a_{vv}(\hat{\mathbf e}_u\hat{\bm\nu}+\hat{\bm\nu}\hat{\mathbf e}_u)}{\sqrt{G_0}}\right]
\\ \nonumber
&+\left[\frac{\partial_vQ_{\nu u}+2a_{vv}Q_{\nu v}}{\sqrt{G_0}}-c_vQ_{uv}\right]\left[\frac{\partial_v(\hat{\mathbf e}_u\hat{\bm\nu}+\hat{\bm\nu}\hat{\mathbf e}_u)}{\sqrt{G_0}}-\frac{a_{uu}(\hat{\mathbf e}_u\hat{\bm\nu}+\hat{\bm\nu}\hat{\mathbf e}_u)}{\sqrt{E_0}}\right]
\\ \nonumber
&+\partial_u\left[\frac{\partial_uQ_{\nu v}+2a_{uu}Q_{\nu u}}{\sqrt{E_0}}-c_uQ_{uv}\right]\frac{\hat{\mathbf e}_v\hat{\bm\nu}+\hat{\bm\nu}\hat{\mathbf e}_v}{\sqrt{E_0}}
+\partial_v\left[\frac{\partial_vQ_{\nu v}-2a_{vv}Q_{\nu u}}{\sqrt{G_0}}+c_v(Q_{uu}+2Q_{\nu\nu})\right]\frac{\hat{\mathbf e}_v\hat{\bm\nu}+\hat{\bm\nu}\hat{\mathbf e}_v}{\sqrt{G_0}}
\\ \nonumber
&+\left[\frac{\partial_uQ_{\nu v}+2a_{uu}Q_{\nu u}}{\sqrt{E_0}}-c_uQ_{uv}\right]\left[\frac{\partial_u(\hat{\mathbf e}_v\hat{\bm\nu}+\hat{\bm\nu}\hat{\mathbf e}_v)}{\sqrt{E_0}}-\frac{a_{vv}(\hat{\mathbf e}_v\hat{\bm\nu}+\hat{\bm\nu}\hat{\mathbf e}_v)}{\sqrt{G_0}}\right]
\\ \nonumber
&+\left[\frac{\partial_vQ_{\nu v}-2a_{vv}Q_{\nu u}}{\sqrt{G_0}}+c_v(Q_{uu}+2Q_{\nu\nu})\right]\left[\frac{\partial_v(\hat{\mathbf e}_v\hat{\bm\nu}+\hat{\bm\nu}\hat{\mathbf e}_v)}{\sqrt{G_0}}-\frac{a_{uu}(\hat{\mathbf e}_v\hat{\bm\nu}+\hat{\bm\nu}\hat{\mathbf e}_v)}{\sqrt{E_0}}\right]
\\ \nonumber
&+\partial_u\left[\frac{\partial_uQ_{\nu\nu}}{\sqrt{E_0}}-2c_uQ_{\nu u}\right]\frac{\hat{\bm\nu}\hat{\bm\nu}}{\sqrt{E_0}}
+\partial_v\left[\frac{\partial_vQ_{\nu\nu}}{\sqrt{G_0}}-2c_vQ_{\nu v}\right]\frac{\hat{\bm\nu}\hat{\bm\nu}}{\sqrt{G_0}}
\\ \nonumber
&+\left[\frac{\partial_uQ_{\nu\nu}}{\sqrt{E_0}}-2c_uQ_{\nu u}\right]\left[\frac{\partial_u(\hat{\bm\nu}\hat{\bm\nu})}{\sqrt{E_0}}-\frac{a_{vv}\hat{\bm\nu}\hat{\bm\nu}}{\sqrt{G_0}}\right]
+\left[\frac{\partial_vQ_{\nu\nu}}{\sqrt{G_0}}-2c_vQ_{\nu v}\right]\left[\frac{\partial_v(\hat{\bm\nu}\hat{\bm\nu})}{\sqrt{G_0}}-\frac{a_{uu}\hat{\bm\nu}\hat{\bm\nu}}{\sqrt{E_0}}\right]
\\ \nonumber
&+c_u\partial_{n}\mathbf Q^{(1)}+c_v\partial_{n}\mathbf Q^{(1)}+\partial_{n}^2\mathbf Q^{(2)}
\end{align}
Finally, the co-rotational derivative of the nematic tensor is
\begin{align}
\bm\omega^{(0)}\cdot\mathbf Q^{(0)}-\mathbf Q^{(0)}\cdot\bm\omega^{(0)}&=
2\omega_{uv}Q_{uv}(\hat{\mathbf e}_u\hat{\mathbf e}_u-\hat{\mathbf e}_v\hat{\mathbf e}_v)
+2\omega_{\nu u}Q_{\nu u}(\hat{\bm\nu}\hat{\bm\nu}-\hat{\mathbf e}_u\hat{\mathbf e}_u)
+2\omega_{\nu v}Q_{\nu v}(\hat{\bm\nu}\hat{\bm\nu}-\hat{\mathbf e}_v\hat{\mathbf e}_v) \\ \nonumber
&-[\omega_{uv}(2Q_{uu}+Q_{\nu\nu})-\omega_{\nu u}Q_{\nu v}-\omega_{\nu v}Q_{\nu u}](\hat{\mathbf e}_u\hat{\mathbf e}_v+\hat{\mathbf e}_v\hat{\mathbf e}_u) \\ \nonumber
&+[\omega_{uv}Q_{\nu v}+\omega_{\nu u}(Q_{uu}-Q_{\nu\nu})+\omega_{\nu v}Q_{uv}](\hat{\mathbf e}_u\hat{\bm\nu}+\hat{\bm\nu}\hat{\mathbf e}_u) \\ \nonumber
&+[-\omega_{uv}Q_{\nu u}+\omega_{\nu v}(Q_{vv}-Q_{\nu\nu})+\omega_{\nu u}Q_{uv}](\hat{\mathbf e}_v\hat{\bm\nu}+\hat{\bm\nu}\hat{\mathbf e}_v)
\end{align}

\section{Linear stability analysis on a flat substrate}\label{app:lsa-flat}
For a flat substrate with $\mathbf v=\bm 0+\delta\mathbf v$, $H=H_0+\delta H$, $\mathbf Q=\mathbf Q_{\rm tangent}+\delta\mathbf Q$, one can use a scaling $\gamma\sim\epsilon^{-1}|\alpha|L$ so that $H/L_a\sim\mathcal{O}(\epsilon^2)$ in the thin film expansion. The pressure at least order becomes $P^{(0)}=2\eta\partial_nv_{\nu}^{(1)}+\alpha Q_{\nu\nu}^{(0)}-\gamma\Delta H$. Then for a flat substrate, the Laplace term $\gamma H\bm\nabla_{\mathcal S}(c_u+c_v)$ in tangential force balance, Eq.~\ref{eq:ftangent}, is replaced by $-\gamma H\bm\nabla_{\mathcal S}(\Delta H)$.

In the deep nematic phase, one assumes perfect alignment $S=1$ with $u_u=\cos(\psi)$, $u_v=\sin(\psi)$ and $u_{\nu}=0$. This gives $\mathbf Q_{\rm tangent}=\frac{1}{4}[1+3\cos(2\psi)]\hat{\mathbf e}_u\hat{\mathbf e}_u+\frac{1}{4}[1-3\cos(2\psi)]\hat{\mathbf e}_v\hat{\mathbf e}_v+\frac{3}{4}\sin(2\psi)(\hat{\mathbf e}_u\hat{\mathbf e}_v+\hat{\mathbf e}_v\hat{\mathbf e}_u)-\frac{1}{2}\hat{\bm\nu}\hat{\bm\nu}$, or $\delta\mathbf Q=\frac{3}{2}\delta\psi[\sin(2\psi)(\hat{\mathbf e}_v\hat{\mathbf e}_v-\hat{\mathbf e}_u\hat{\mathbf e}_u)+\cos(2\psi)(\hat{\mathbf e}_u\hat{\mathbf e}_v+\hat{\mathbf e}_v\hat{\mathbf e}_u)]$.

Arbitrarily choosing $\psi=0$ such that $Q_{uu}=1$, $Q_{vv}=-1/2$, $Q_{uv}=Q_{\nu i}=0$, one gets $\delta\mathbf Q=\frac{3}{2}\delta\psi(\hat{\mathbf e}_u\hat{\mathbf e}_v+\hat{\mathbf e}_v\hat{\mathbf e}_u)$. Expanding the field perturbations in Fourier space with wavevector $\mathbf q$, all fields follow $\delta\varphi(\mathbf r,t)=\hat{\varphi}(t)\exp[\mathrm{i}\mathbf q\cdot\mathbf{r}]$. One gets from orientation dynamics and tangential force balance 
\begin{align}
\Gamma\partial_t\hat{\psi}+Kq^2\hat{\psi}
&=\frac{\mathrm{i}\Gamma}{2}(\lambda+1)q_u\hat{v}_v+\frac{\mathrm{i}\Gamma}{2}(\lambda-1)q_v\hat{v}_u \\
\mathrm{i}\alpha[3(1-\delta_{ij})q_j\hat{\psi}+(2Q_{ij}q_j+q_i)\hat{H}/H_0]-2\mathrm{i}\gamma H_0 q_i q^2\hat{H}/H_0
&=2[(\xi_s/H_0+\eta q^2)\delta_{ij}+3\eta q_iq_j]\hat{v}_j
\end{align}
or explicitly in components
\begin{align}
3\mathrm{i}\alpha q_v\hat{\psi}+\mathrm{i}q_u[3\alpha-2\gamma H_0q^2]\hat{H}/H_0
&=2(\xi_s/H_0+\eta q^2+3\eta q_u^2)\hat{v}_u+6\eta q_uq_v\hat{v}_v \\
3\mathrm{i}\alpha\,q_u\hat{\psi}-2\mathrm{i}\gamma H_0q_vq^2\hat{H}/H_0
&=2(\xi_s/H_0+\eta q^2+3\eta q_v^2)\hat{v}_v+6\eta q_uq_v\hat{v}_u
\end{align}
These equations are inverted in terms of $\hat{v}_i$ and the evolution equation of $\hat{\psi}$ depends on $\hat\psi$ and $\hat H$. The evolution equation for $\hat H$ follows
\begin{equation}
\partial_t\frac{\hat{H}}{H_0}=\frac{3\alpha q_uq_v}{\xi_s/H_0+4\eta q^2}\hat{\psi}-\frac{1}{2}\frac{2\gamma H_0 q^4-3\alpha q_u^2}{\xi_s/H_0+4\eta q^2}\frac{\hat{H}}{H_0}
\end{equation}
Thus in general, one obtains coupled equations between perturbations of orientation $\hat{\psi}$ and thickness $\hat{H}$.

We call $\phi$ the angle of the wavevector $\mathbf q$ with respect to the initial nematic orientation, $\bm q=q(\cos\phi\,\hat{\mathbf e}_u+\sin\phi\,\hat{\mathbf e}_v)$. We consider separately longitudinal ($\phi=0$) and transverse ($\phi=\pi/2$) fluctuations, which allows a decoupling of the dynamic equations in Fourier space. For transverse fluctuations $q_u=0$, one finds stability of a uniform thickness, with standard orientational instability for extensile activity ($\partial_t\hat{\psi}\sim-\alpha(\lambda-1)\hat{\psi}$) assuming $\lambda>1$. For longitudinal fluctuations $q_v=0$, one gets
\begin{align}
\left[\partial_t+\frac{K q^2}{\Gamma}\right]\hat{\psi}&=-\frac{3\alpha(\lambda+1)q^2}{4(\xi_s/H_0+\eta q^2)}\hat{\psi} \\
\partial_t\frac{\hat{H}}{H_0}&=\frac{q^2}{2}\frac{-2\gamma H_0 q^2+3\alpha}{\xi_s/H_0+4\eta q^2}\frac{\hat{H}}{H_0}
\end{align}
Thus, thickness growth occurs for contractile activity $\alpha>0$, and inhibits orientational fluctuations when $\lambda>0$. The existence of a non-zero surface tension sets a characteristic length-scale $\tilde{\lambda}_a=\sqrt{\gamma H_0/\alpha}$ for the longitudinal instability to develop.

Until now, we assumed the nematic tensor to remain perfectly tangential to the substrate. However as discussed in the main text, making a perturbation of the out-of-plane components provides
\begin{align}
\partial_t\delta Q_{\nu u}-\frac{K}{\Gamma}\Delta\delta Q_{\nu u}&=-\left[\frac{3\alpha}{4\eta}(\lambda-1)+\frac{w}{2\Gamma H_0}\right]\delta Q_{\nu u} \\
\partial_t\delta Q_{\nu v}-\frac{K}{\Gamma}\Delta\delta Q_{\nu v}&=-\left[\frac{3\lambda\alpha}{4\eta}+\frac{w}{2\Gamma H_0}\right]\delta Q_{\nu v} \\
\partial_t\delta Q_{\nu\nu}-\frac{K}{\Gamma}\Delta\delta Q_{\nu\nu}&=\frac{3\lambda}{2}\partial_t\frac{\hat{H}}{H_0}-\frac{2w}{3\Gamma H_0}\delta Q_{\nu\nu}
\end{align}
Thus for $\alpha,\lambda>0$, the components $\delta Q_{\nu i}$ vanish but the thickness instability triggers variations of $\delta Q_{\nu\nu}$ for $w\lesssim\Gamma H_0\alpha/\eta$. For extensile activity $\alpha<0$, thickness variations vanish but the components $\delta Q_{\nu i}$ are destabilized.

\section{Linear stability analysis on a sinusoidal substrate}\label{app:lsa-sin}

Here we perform linear stability analysis for an active nematic shell on a sinusoidal substrate with Cartesian coordinates $x$, $y=v$ and $z=c_0L^2\cos(2\pi\,u/L)$, with $u=[0,L]$ the arc-length and $c_0$ the characteristic substrate curvature. We assume invariance along $y$ in the following. One finds the orthogonal basis vectors $\mathbf b_u=\hat{\mathbf e}_x-(2\pi c_0L)\sin(2\pi\,u/L)\hat{\mathbf e}_z$, $\mathbf b_v=\hat{\mathbf e}_y$, $\hat{\bm\nu}=[\hat{\mathbf e}_z+(2\pi c_0L)\sin(2\pi\,u/L)\hat{\mathbf e}_x]/\sqrt{E_0}$ with $E_0=1+(2\pi c_0L)^2\sin^2(2\pi\,u/L)$. One also has the identities $a_{uu}=a_{vv}=c_v=0$, $G_0=1$, and substrate curvature
\begin{align}
c_u(u)=4\pi^2\,c_0\cos(2\pi\,u/L)\,E_0^{-3/2}
\end{align}
Thus, the non-trivial basis vector derivatives are $\partial_u\hat{\mathbf e}_u=-c_u\sqrt{E_0}\hat{\bm\nu}$ and $\partial_u\hat{\bm\nu}=c_u\sqrt{E_0}\hat{\mathbf e}_u$. Using $E_0(n)=E_0[1+nc_u]^2$ The gradient tensor reads $\bm\nabla=[E_0(n)]^{-1/2}\hat{\mathbf e}_u\partial_u+\hat{\bm\nu}\partial_n$ for $\partial_v=0$.

The incompressibility equation reads $\partial_u v_u/\sqrt{E_0}+\partial_n v_{\nu}=0$ with $\bm\nabla_{\mathcal S}\cdot\mathbf v_{\mathcal S}=\partial_u v_u/\sqrt{E_0}$, the symmetric strain rate is
$\tilde{\mathbf u}=(\partial_uv_u/\sqrt{E_0})(\hat{\mathbf e}_u\hat{\mathbf e}_u-\hat{\bm\nu}\hat{\bm\nu})+\frac{1}{2}(\partial_uv_v/\sqrt{E_0})(\hat{\mathbf e}_u\hat{\mathbf e}_v+\hat{\mathbf e}_v\hat{\mathbf e}_u)-\frac{\alpha}{2\eta}Q_{\nu i}(\hat{\mathbf e}_i\hat{\bm\nu}+\hat{\bm\nu}\hat{\mathbf e}_i)$ where one used $\partial_nv_i^{(1)}=c_iv_i-\alpha Q_{\nu i}/\eta$. The vorticity becomes $\bm\omega=\frac{1}{2}(\partial_uv_v/\sqrt{E_0})(\hat{\mathbf e}_u\hat{\mathbf e}_v-\hat{\mathbf e}_v\hat{\mathbf e}_u)+(c_iv_i-\alpha Q_{\nu i}/2\eta)(\hat{\bm\nu}\hat{\mathbf e}_i-\hat{\mathbf e}_i\hat{\bm\nu})$. The projected Laplacian of the velocity field reads $\bm 1_{\mathcal S}\cdot\Delta\mathbf v=E_0^{-1}\partial_u^2v_u\hat{\mathbf e}_u+\partial_u^2v_v\hat{\mathbf e}_v-c_u^2v_u\hat{\mathbf e}_u$.

The nematic tensor gradient is
\begin{align}
&[\bm\nabla\mathbf Q]^{(0)}=
\hat{\mathbf e}_u\left[\frac{\partial_uQ_{uu}}{\sqrt{E_0}}+2c_uQ_{\nu u}\right]\hat{\mathbf e}_u\hat{\mathbf e}_u
+\hat{\mathbf e}_u\left[\frac{\partial_uQ_{uv}}{\sqrt{E_0}}+c_uQ_{\nu v}\right](\hat{\mathbf e}_u\hat{\mathbf e}_v+\hat{\mathbf e}_v\hat{\mathbf e}_u)
-\hat{\mathbf e}_u\left[\frac{\partial_uQ_{uu}+\partial_uQ_{\nu\nu}}{\sqrt{E_0}}\right]\hat{\mathbf e}_v\hat{\mathbf e}_v
\\ \nonumber
&+\hat{\mathbf e}_u\left[\frac{\partial_uQ_{\nu u}}{\sqrt{E_0}}+c_u(Q_{\nu\nu}-Q_{uu})\right](\hat{\mathbf e}_u\hat{\bm\nu}+\hat{\bm\nu}\hat{\mathbf e}_u)
+\hat{\mathbf e}_u\left[\frac{\partial_uQ_{\nu v}}{\sqrt{E_0}}-c_uQ_{uv}\right](\hat{\mathbf e}_v\hat{\bm\nu}+\hat{\bm\nu}\hat{\mathbf e}_v)
+\hat{\mathbf e}_u\left[\frac{\partial_uQ_{\nu\nu}}{\sqrt{E_0}}-2c_uQ_{\nu u}\right]\hat{\bm\nu}\hat{\bm\nu}
\end{align}
The nematic divergence becomes $\bm\nabla\cdot\mathbf Q=[\partial_uQ_{uu}/\sqrt{E_0}+2c_uQ_{\nu u}]\hat{\mathbf e}_u+[\partial_uQ_{uv}/\sqrt{E_0}+c_uQ_{\nu v}]\hat{\mathbf e}_v+[\partial_uQ_{\nu u}/\sqrt{E_0}+c_u(Q_{\nu\nu}-Q_{uu})]\hat{\bm\nu}$. Finally, the nematic Laplacian reads
\begin{align}
&\Delta\mathbf Q^{(0)}= \\ \nonumber
&+\partial_u\left(\frac{\partial_uQ_{uu}}{\sqrt{E_0}}\right)
\frac{\hat{\mathbf e}_u\hat{\mathbf e}_u}{\sqrt{E_0}}
-\partial_u\left(\frac{\partial_uQ_{uu}+\partial_uQ_{\nu\nu}}{\sqrt{E_0}}\right)\frac{\hat{\mathbf e}_v\hat{\mathbf e}_v}{\sqrt{E_0}}
+\partial_u\left(\frac{\partial_uQ_{\nu\nu}}{\sqrt{E_0}}\right)
\frac{\hat{\bm\nu}\hat{\bm\nu}}{\sqrt{E_0}}
\\ \nonumber
&+2c_u\left[\partial_uQ_{\nu u}/\sqrt{E_0}+c_u(Q_{\nu\nu}-Q_{uu})+Q_{\nu u}/\sqrt{E_0}\right]
(\hat{\mathbf e}_u\hat{\mathbf e}_u-\hat{\bm\nu}\hat{\bm\nu})
\\ \nonumber
&+\partial_u\left[\partial_uQ_{uv}/\sqrt{E_0}+c_uQ_{\nu v}\right]
\frac{\hat{\mathbf e}_u\hat{\mathbf e}_v+\hat{\mathbf e}_v\hat{\mathbf e}_u}{\sqrt{E_0}}
+c_u\left[\partial_uQ_{\nu v}/\sqrt{E_0}-c_uQ_{uv}\right]
(\hat{\mathbf e}_u\hat{\mathbf e}_v+\hat{\mathbf e}_v\hat{\mathbf e}_u)
\\ \nonumber
&+\partial_u\left[\partial_uQ_{\nu u}/\sqrt{E_0}+c_u(Q_{\nu\nu}-Q_{uu})\right]\frac{\hat{\mathbf e}_u\hat{\bm\nu}+\hat{\bm\nu}\hat{\mathbf e}_u}{\sqrt{E_0}}
+c_u\left[\partial_u(Q_{\nu\nu}-Q_{uu})/\sqrt{E_0}-4c_uQ_{\nu u}\right]
(\hat{\mathbf e}_u\hat{\bm\nu}+\hat{\bm\nu}\hat{\mathbf e}_u)
\\ \nonumber
&+\partial_u\left[\partial_uQ_{\nu v}/\sqrt{E_0}-c_uQ_{uv}\right]
\frac{\hat{\mathbf e}_v\hat{\bm\nu}+\hat{\bm\nu}\hat{\mathbf e}_v}{\sqrt{E_0}}
-c_u\left[\partial_uQ_{uv}/\sqrt{E_0}+c_uQ_{\nu v}\right]
(\hat{\bm\nu}\hat{\mathbf e}_v+\hat{\mathbf e}_v\hat{\bm\nu})
\end{align}

Similarly to Ref.~\cite{Bell2022}, we study how curvature impacts the spontaneous flow transition driven by active stresses on a sinusoidal substrate. We consider a static reference state $\mathbf v=\bm 0+\delta\mathbf v$, $\mathbf Q=\mathbf Q_0+\delta\mathbf Q$, $H=H_0+\delta H$. Note that the substrate has spatial gradients of curvature, and the reference state can be static only if surface tension $\gamma$ vanishes, which we assume in the following. We take the limit of infinite tangential anchoring for compatibility with the case $K_c\neq 0$, such that $Q_{\nu\nu}+1/2=Q_{\nu i}=0$. We also consider a deep nematic phase $\chi\rightarrow\infty$ and write $\mathbf Q=\frac{1}{4}[1+3\cos(2\psi)]\hat{\mathbf e}_u\hat{\mathbf e}_u+\frac{1}{4}[1-3\cos(2\psi)]\hat{\mathbf e}_v\hat{\mathbf e}_v+\frac{3}{4}\sin(2\psi)(\hat{\mathbf e}_u\hat{\mathbf e}_v+\hat{\mathbf e}_v\hat{\mathbf e}_u)-\frac{1}{2}\hat{\bm\nu}\hat{\bm\nu}$ with the director angle $\psi=\psi_0+\delta\psi$ and $\delta\mathbf Q=\frac{3}{2}\delta\psi[\sin(2\psi_0)(\hat{\mathbf e}_v\hat{\mathbf e}_v-\hat{\mathbf e}_u\hat{\mathbf e}_u)+\cos(2\psi_0)(\hat{\mathbf e}_u\hat{\mathbf e}_v+\hat{\mathbf e}_v\hat{\mathbf e}_u)]$.
\\\\
First, we are interested in a uniform orientation along $\hat{\mathbf e}_v$ (longitudinal) with strong anchoring at $u=0,L$, such that $\psi_0=\pi/2$, $\mathbf Q_0=\hat{\mathbf e}_v\hat{\mathbf e}_v-\frac{1}{2}(\hat{\mathbf e}_u\hat{\mathbf e}_u+\hat{\bm\nu}\hat{\bm\nu})$ and $\delta\mathbf Q=-\frac{3}{2}\delta\psi(\hat{\mathbf e}_u\hat{\mathbf e}_v+\hat{\mathbf e}_v\hat{\mathbf e}_u)$.
The perturbation equations give
\begin{align}
\Gamma\partial_t\delta\psi
&=\frac{K}{\sqrt{E_0}}\partial_u\left[\frac{\partial_u\delta\psi}{\sqrt{E_0}}\right]-Kc_u^2\delta\psi
-\frac{K_cc_u}{2H_0}\delta\psi-\frac{\Gamma}{2\sqrt{E_0}}(\lambda-1)\partial_u\delta v_v \\
\delta v_u&=\delta H=0 \\
-\frac{3\alpha H_0}{2\sqrt{E_0}}\partial_u\delta\psi
&=\xi_s\delta v_v-\eta H_0\partial_u^2\delta v_v
\end{align}
To remove geometric non-linearities, we assume $(c_0L)^2\ll 1$ such that $E_0\simeq 1$.
We perform the expansions $\delta\psi(u,t)=\sum_k\,\psi_k(t)\sin(\pi k u/L)$, $\delta v_v(u,t)=\sum_k\,v_k(t)\cos(\pi k u/L)$ which satisfy the anchoring conditions $\delta\psi(0,L)=0$. 
We end up with the first two modes
\begin{align}
\Gamma\partial_t\psi_1&=
-\frac{\pi^2K}{L^2}\psi_1-\frac{3\pi^2\alpha H_0(\lambda-1)\Gamma}{4L^2(\xi_s+\eta H_0(\pi/L)^2)}\,\psi_1
+\frac{\pi^2K_cc_0}{H_0}\psi_1 \\
\Gamma\partial_t\psi_2&=-\frac{\pi^2K}{L^2}\psi_2-\frac{3\pi^2\alpha H_0(\lambda-1)\Gamma}{L^2(\xi_s+\eta H_0(2\pi/L)^2)}\psi_2
\end{align}
It shows the usual spontaneous flow transition for extensile activity $\alpha<0$ and rod-like flow alignment $\lambda>0$. The curvature-orientation coupling also triggers an instability when $K_cc_0>0$, with steady shear flow for non-zero activity.
\\\\
Then we consider a transverse orientation along $\hat{\mathbf e}_u$ such that $\psi_0=0$, $\mathbf Q_0=\hat{\mathbf e}_u\hat{\mathbf e}_u-\frac{1}{2}(\hat{\mathbf e}_v\hat{\mathbf e}_v+\hat{\bm\nu}\hat{\bm\nu})$ and $\delta\mathbf Q=\frac{3}{2}\delta\psi(\hat{\mathbf e}_u\hat{\mathbf e}_v+\hat{\mathbf e}_v\hat{\mathbf e}_u)$. One ends up with the following system of equations
\begin{align}
\Gamma\partial_t\delta\psi&=K[\partial_u^2\delta\psi-c_u^2\delta\psi]+\frac{\Gamma}{2}(\lambda+1)\partial_u\delta v_v-\frac{K_cc_u}{2H_0}\delta\psi \\
\partial_t\delta H&=-H_0\partial_u\delta v_u \\
\frac{3}{2}\alpha\partial_u\delta H&=\xi_s\delta v_u-4\eta H_0\partial_u^2\delta v_u \\
\frac{3}{2}\alpha H_0\,\partial_u\delta\psi&=\xi_s\delta v_v-\eta H_0\partial_u^2\delta v_v
\end{align}
Performing again a trigonometric expansion in the fields with $\delta\psi(0,L)=0$, one finds
\begin{align}
\Gamma\partial_t\psi_1&=
-\frac{\pi^2K}{L^2}\psi_1-\frac{3\pi^2\alpha H_0(\lambda+1)\Gamma}{4L^2(\xi_s+\eta H_0(\pi/L)^2)}\,\psi_1
+\frac{\pi^2K_cc_0}{H_0}\psi_1 \\
\partial_tH_k&=\frac{3\pi^2\alpha H_0}{2L^2[\xi_s+4\eta H_0 (\pi k/L)^2]}k^2H_k
\end{align}
On top of the usual flow transition for extensile activity ($\alpha<0$ and $\lambda>0$), one finds a similar destabilization from the curvature-orientation coupling when $K_cc_0>0$. In addition, this transverse mode allows for a coupling between transverse velocity $v_u$ and thickness $H$, such that an instability develops for contractile activity ($\alpha>0$). Then, surface tension $\gamma$ must be non-zero to stabilize a finite steady-state.

\end{widetext}

%



\end{document}